\documentclass[usenatbib]{mnras}
\usepackage{natbib,amssymb,amsmath,times,graphicx,url,aas_macros}
\usepackage{setspace}      % To manually set line spacing
\usepackage{epstopdf}       % to convert eps images to pdf images
\usepackage{verbatim}       % to allow multiline comments
\usepackage{bm}		  % allow shorthand to bold text
\usepackage{fleqn}		  % flush left equations
\usepackage{Wurster_macros}  % my macros

\title[On the origin of magnetic fields in stars]{On the origin of magnetic fields in stars}

\author[Wurster, Bate \& Price]{James Wurster$^{1}$\thanks{j.wurster@exeter.ac.uk}, Matthew R. Bate$^{1}$\thanks{mbate@astro.ex.ac.uk}, and Daniel J. Price$^{2}$ \\
$^{1}$School of Physics and Astronomy, University of Exeter, Stocker Rd, Exeter EX4 4QL, UK \\
$^{2}$Monash Centre for Astrophysics and School of Physics and Astronomy, Monash University, Vic 3800, Australia \\
}
\date{Submitted: Revised: Accepted: }
\pagerange{\pageref{firstpage}--\pageref{lastpage}} \pubyear{2018}

\begin{document}
\label{firstpage}
\bibliographystyle{mnras}
\maketitle

\begin{abstract}
 Are the kG-strength magnetic fields observed in young stars a fossil field left over from their formation or are they generated by a dynamo? We use radiation non-ideal magnetohydrodynamics simulations of the gravitational collapse of a rotating, magnetized molecular cloud core over 17 orders of magnitude in density, past the first hydrostatic core to the formation of the second, stellar core, to examine the fossil field hypothesis.  Whereas in previous work we found that magnetic fields in excess of 10 kG can be implanted in stars at birth, this assumed ideal magnetohydrodynamics (MHD), i.e. that the gas is coupled to the magnetic field. Here we present non-ideal MHD calculations which include Ohmic resistivity, ambipolar diffusion and the Hall effect. For realistic cosmic ray ionization rates, we find that magnetic field strengths of $\lesssim$ kG are implanted in the stellar core at birth, ruling out a strong fossil field.  While these results remain sensitive to resolution, they cautiously provide evidence against a fossil field origin for stellar magnetic fields, suggesting instead that magnetic fields in stars originate in a dynamo process.
 \end{abstract}

%---------------------------------------------------------------------------------------------------------------
\section{Introduction}
\label{intro}

All low-mass stars are magnetized, including our Sun, but the origin of stellar magnetic fields is uncertain. Low-mass stars tend to have strong, kG-strength surface magnetic field strengths when they are young that weaken as they age. This long-term evolution is consistent with their magnetic fields being generated by convective dynamos, because stellar rotation rates also decrease with time due to the emission of magnetized winds and outflows \citep[e.g.][]{Parker1958,Schatzman1962,WebDav1967,Skumanich1972,PizMagMicSciVen2003,WriDraMamHen2011,VidottoEtAl2014,SeeEtAl2015}.  In addition, since young, low-mass stars are fully convective, it is generally assumed that any birth magnetic fields are quickly diffused and replaced by dynamo-generated fields \citep{ChabrierKuker2006}.  However, there is observed to be a large dispersion in the magnetic field strengths of young stars \citep[e.g.][]{JohnsKrull2007,YanJoh2011}, and these studies have so far failed to find any correlation between the measured magnetic field properties and the stellar properties thought to be important for dynamo action.  This has lead to speculation that the magnetic fields of low-mass stars may be dominated by primordial or `fossil' magnetic fields that are implanted during the star formation process  \citep{Tayler1987,Moss2003,TouWicFer2004,YanJoh2011}.  
However, the strength and geometry of magnetic fields implanted in protostars during the star formation process is unknown. 

The first numerical studies to model the collapse of a magnetized molecular cloud all the way to stellar core formation were performed by \citet{TomidaEtAl2013} and \citet*{BatTriPri2014}.  \citet*{BatTriPri2014} showed that magnetic fields with strengths in excess of 10~kG may be implanted in the stellar core at birth.  This supported the hypothesis that the strong magnetic fields observed in young low-mass stars may be fossil fields.  However, these calculations employed the ideal magnetohydrodynamics (MHD) approximation, whereby the gas was assumed to be sufficiently ionized such that the magnetic field was `frozen into' the fluid as it collapses.

The molecular clouds where stars are born are only weakly ionized \citep[e.g.][]{MesSpi1956,NakUme1986,UmeNak1990}, implying that  the ideal MHD approximation is not valid.  Weakly ionized gas gives rise to three main non-ideal MHD effects -- ambipolar (ion-neutral) diffusion, Ohmic resistivity and the Hall effect -- where the relative importance of each of these depends, amongst other things, on the gas density and temperature, and magnetic field strength \citep[e.g.][]{WarNg1999, NakNisUme2002, TasMou2007b, Wardle2007, PanWar2008, KeiWar2014}.  Ambipolar diffusion and Ohmic resistivity lead to the diffusion of gas relative to the magnetic field and, therefore, are likely to lead to weaker fossil magnetic fields.  The Hall effect is not diffusive but modifies the geometry of the magnetic field to increase or decrease the angular momentum in the dense gas that collapses to the equatorial plane \citep{BraWar2012}.  Recent star formation studies have included some or all of these non-ideal effects \citep[e.g.][]{TomidaEtAl2013,TomOkuMac2015,TsukamotoEtAl2015_oa,TsukamotoEtAl2015_hall,TsukamotoEtAl2017,Masson+2016,WurPriBat2016,WurPriBat2017,WurBatPri2018,VaytetEtAl2018}.  

In this paper, we build on \citet*{BatTriPri2014} and \citet*{WurBatPri2018} by modelling the gravitational collapse of a molecular cloud core through the first and stellar core phases to determine the magnetic field strength implanted in the stellar core.  In our primary analysis, we compare an ideal MHD model to a non-ideal MHD model that includes a self-consistent treatment of the non-ideal processes.  In Section~\ref{sec:methods}, we summarise our methods and in Section~\ref{sec:ic} we present our initial conditions.  Our results are presented in Section~\ref{sec:results}, we discuss the caveats in Section~\ref{sec:dis}, and we conclude in Section~\ref{sec:conc}.

%----------------------------------------------------------------------------------------------------------------
\section{Methods}
\label{sec:methods}

Our method is almost identical to that employed by \citet*{WurBatPri2018}:  To solve the self-gravitating, radiation non-ideal magnetohydrodynamics equations, we use \textsc{sphNG}, which is a three-dimensional Lagrangian smoothed particle hydrodynamics (SPH) code that originated from \citet{Benz1990} but has been substantially extended over the past 30 years to include (e.g.) a consistent treatment of variable smoothing lengths \citep{PriMon2007}, individual timestepping \citep{BatBonPri1995}, radiation as flux limited diffusion \citep{WhiBatMon2005,WhiBat2006}, magnetic fields \citep[for a review, see][]{Price2012}, and non-ideal MHD \citep{WurPriAyl2014,WurPriBat2016}.  For stability of the magnetic field, we use the source-term subtraction approach \citep{BorOmaTru2001}, constrained hyperbolic/parabolic divergence cleaning \citep{TriPri2012,TriPriBat2016}, and artificial resistivity as described in \cite{Phantom2017}.

We use Version 1.2.1 of the \textsc{Nicil} library \citep{Wurster2016} to self-consistently calculate the non-ideal MHD coefficients, using the canonical cosmic ray ionization rate of \zetaeq{-17}  (\citealp{SpiTom1968}; important at low temperatures and densities) and thermal ionization (important at high temperatures and densities).  We include three non-ideal MHD terms: Ohmic resistivity, ambipolar diffusion and the Hall effect.

%----------------------------------------------------------------------------------------------------------------
\section{Initial conditions}
\label{sec:ic}

Our initial conditions are identical to those in  \citet*{BatTriPri2014} and \citet*{WurBatPri2018}:  A 1~M$_{\odot}$ slowly rotating spherical molecular cloud core of uniform density is placed in pressure equilibrium with a warm, low-density ambient medium. The core has radius $R_\text{c} = 4\times10^{16}$~cm, an initial (isothermal) sound speed of $c_\text{s} = \sqrt{p/\rho}= 2.2\times 10^{4}$~cm~s$^{-1}$, and a solid body rotation about the $z$-axis with $\Omega = 1.77 \times 10^{-13}$~rad s$^{-1}$; this rotation corresponds to a ratio of rotational to gravitational energy $\beta_\text{r} \simeq 0.005$. The magnetic field is initially uniform in the $z$-direction, parallel to and aligned with the rotation axis.  The initial magnetic field strength is $B_0 = 163\mu$G, which is equivalent to a mass-to-flux ratio of $\mu_0\equiv \mu(R_\text{c}) = 5$ in units of the critical mass-to-flux ratio \citep[e.g.][]{Mestel1999,MacKle2004}.  We define
\begin{flalign}
\label{eq:masstofluxmu}
\mu(r) &\equiv \frac{M/\Phi_\text{B}}{\left(M/\Phi_\text{B}\right)_\text{crit}},&
\end{flalign}
where
\begin{flalign}
\label{eq:masstoflux}
\frac{M}{\Phi_\text{B}} &\equiv \frac{M(r)}{\pi r^2 B(r)},&
\end{flalign}
is the mass-to-flux ratio and 
\begin{flalign}
\label{eq:masstofluxcrit}
\left(\frac{M}{\Phi_\text{B}}\right)_\text{crit} &= \frac{c_1}{3\pi}\sqrt{ \frac{5}{G} },&
\end{flalign}
is the critical value given in CGS units where the gravitational and magnetic forces balance.  In these equations, $M(r)$ is the total mass contained within a sphere of radius $r$, $G$ is the gravitational constant and $c_1 \simeq 0.53$ is a dimensionless coefficient numerically determined by \citet{MouSpi1976}.  Following \citet{Joos+2013}, we define $B(r)$ to be the average magnetic field strength in a shell at distance $r$.

There are $3 \times 10^{6}$ equal mass SPH particles in the core, and $1.46 \times 10^{6}$ particles in the surrounding medium.  This resolution was found to be adequate to capture the evolution accurately in the ideal MHD calculations of \citet{BatTriPri2014}.
%----------------------------------------------------------------------------------------------------------------
\section{Results}
\label{sec:results}

\begin{figure} 
\centering
\includegraphics[width=\columnwidth]{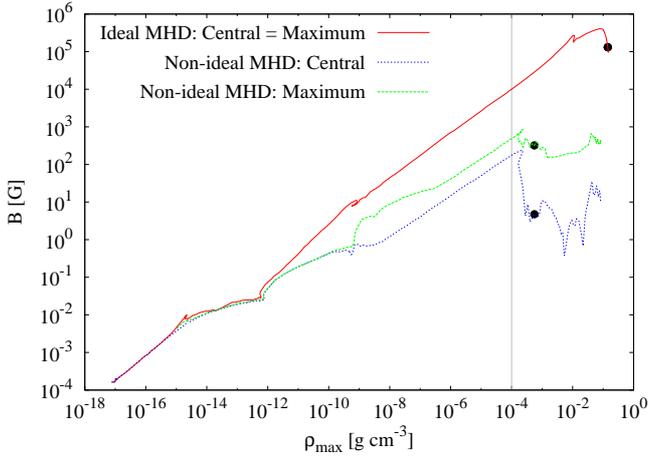}
\caption{Evolution of the magnetic field strength against maximum density, which is a proxy for time.  The vertical grey line indicates the formation density of the stellar core, and the black circles are placed 6 months after the formation of the stellar core for each model (i.e. \dtsc{0.5}).  For both models, the maximum density is coincident with the centre of the system (i.e. $\rho_\text{max} = \rho_\text{cen}$).  In the ideal MHD model, the maximum and central magnetic field strength are the same for the entire simulation.  In the non-ideal MHD model, the central and maximum magnetic field strengths are no longer coincident near the end of the first core phase, and at stellar core formation, are a few orders of magnitude lower than the values in the ideal MHD model.  }
\label{fig:Vrho}
\end{figure}

Fig.~\ref{fig:Vrho} shows the evolution of the central and maximum magnetic field strengths for the ideal and non-ideal MHD models.  During the isothermal collapse and early stages of the first core evolution (\rhoxls{-12}; \citealt{Larson1969}), the central magnetic field strength increases and is independent of the ionization rate.  In this regime, the density is too low and the  magnetic field strength is too weak for the non-ideal effects to change the dynamics of the collapse.  During the first core phase, the magnetic field strength grows rapidly with density for the ideal MHD model ($B_\text{max} \propto \rho_\text{max}^{0.8}$) \citep[as previously seen in, e.g.,][]{BatTriPri2014,TsukamotoEtAl2015_oa,WurBatPri2018}. This is a stronger relation than the $B_\text{max} \propto \rho_\text{max}^{2/3}$ relation expected in isotropic contraction, since the mass is primarily accreted via the equatorial plane.  This new gas drags magnetic field lines with it, amplifying the central strength.  In the non-ideal MHD model, Ohmic resistivity diffuses the magnetic field, thus the first core increases in density without a strong amplification of the central magnetic field strength; this results in the tapered growth rate during this phase.

\begin{figure} 
\centering
\includegraphics[width=\columnwidth,trim={0 2cm 0 0}]{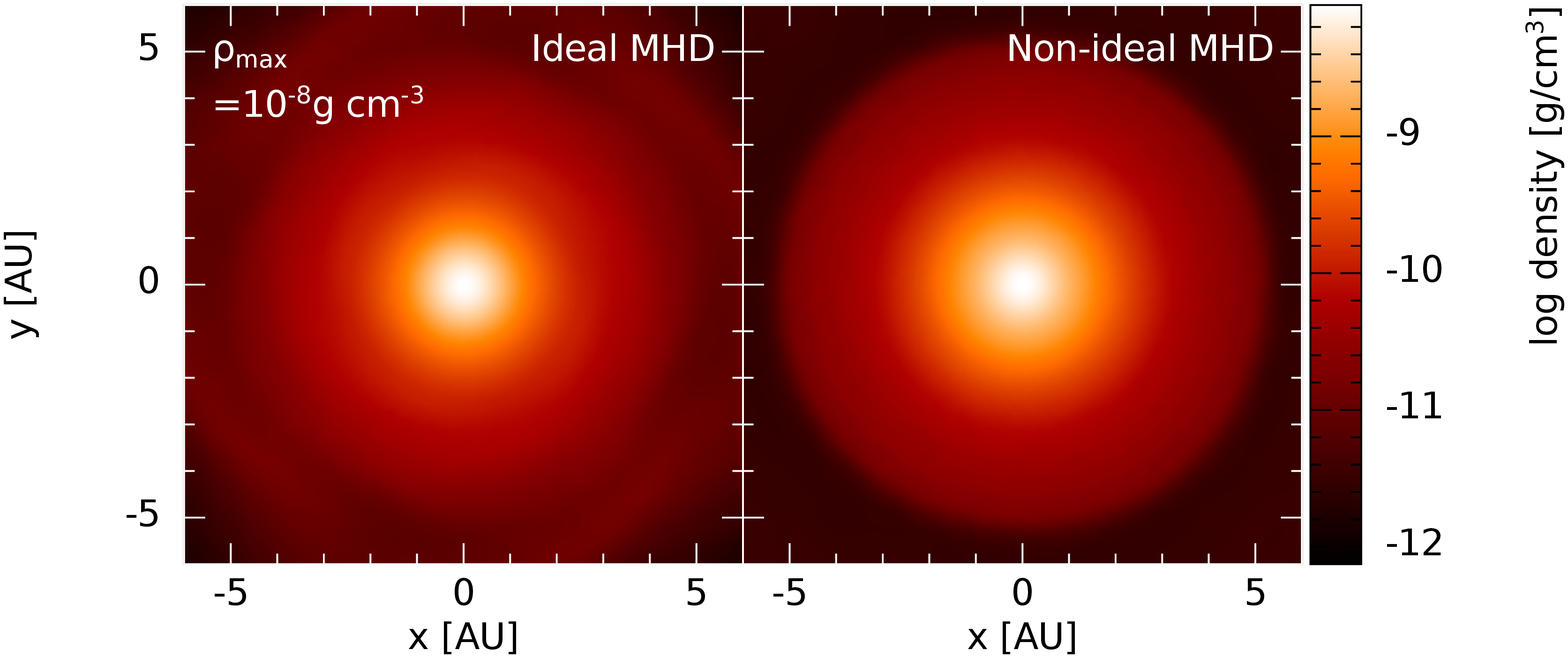}
\includegraphics[width=\columnwidth]{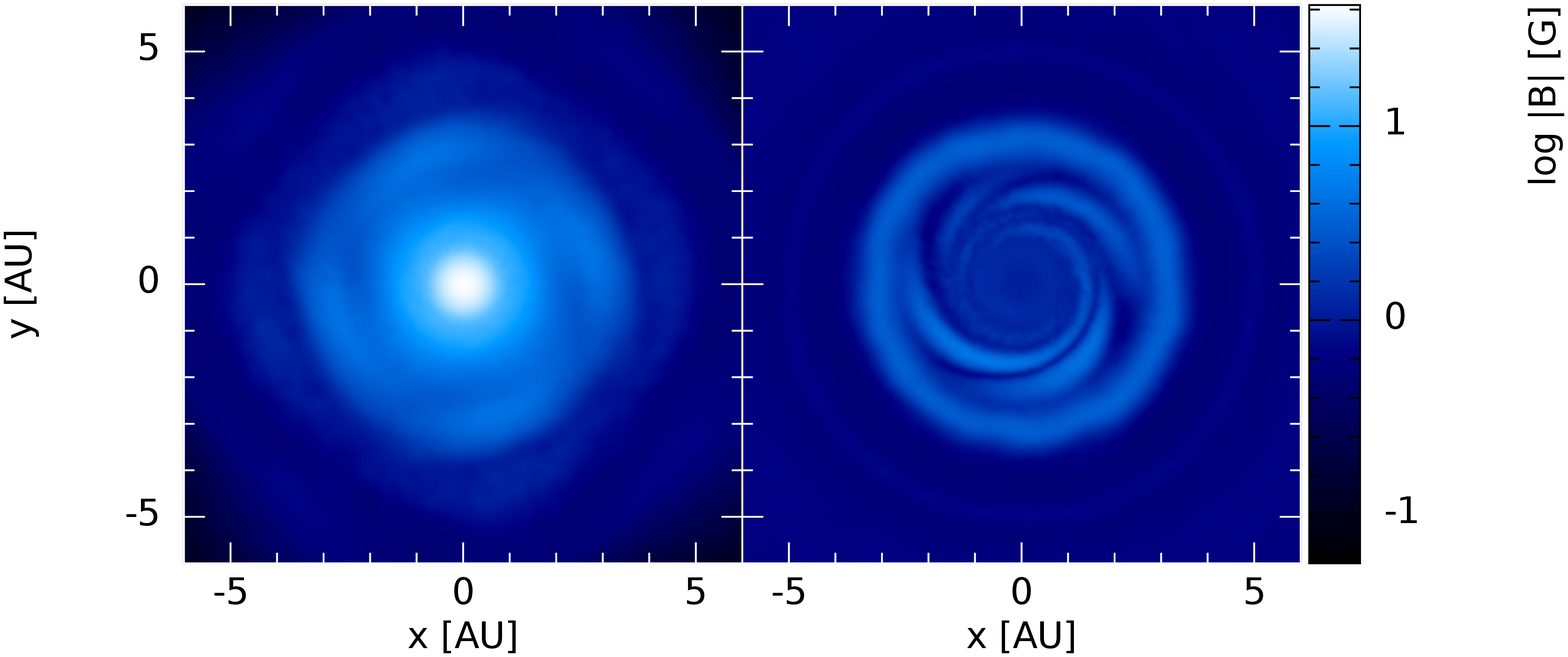}
\caption{Density (top) and magnetic field strength (bottom) slices through the first hydrostatic core perpendicular to the rotation axis for the ideal (left) and non-ideal (right) MHD models once the maximum density has reached $\rho_\text{max} \approx 10^{-8}$~g~cm$^{-3}$ (near the end of the first core phase).  The initial magnetic field strength in both models is five times the critical mass-to-flux ratio.  The density slices are qualitatively similar.  The magnetic field strength follows the density profile in the ideal MHD model, while the maximum magnetic field strength in the non-ideal MHD model is in a structured torus at 1-3~au from the centre of the core.}
\label{fig:rhobfield:fhc}
\end{figure}

By the end of the first core phase (\rhoxapprox{-8}), the maximum magnetic field strengths differ by a factor of $\sim$6 between the two models.  Of even greater importance, the large physical resistivity in the non-ideal MHD model diffuses the magnetic field out of the centre of the first core, leaving the strongest magnetic field to be in a wide, structured torus ranging 1-3~au from the centre of the first core, but still within the first core; see Fig.~\ref{fig:rhobfield:fhc}.  This is the formation of the so-called `magnetic wall' \citep[e.g.][]{LiMck1996,TasMou2005b,TasMou2007b,TasMou2007c}; this feature was discussed in \citet{TomOkuMac2015}, however, the structure was not prominent in their model.  The central magnetic field strengths of the two models differ by a factor of $\sim$35.  Thus, despite the two models having similar density profiles during the first core phase (top row of Fig.~\ref{fig:rhobfield:fhc}), their magnetic field structures differ considerably.

\begin{figure} 
\centering
\includegraphics[width=0.8\columnwidth]{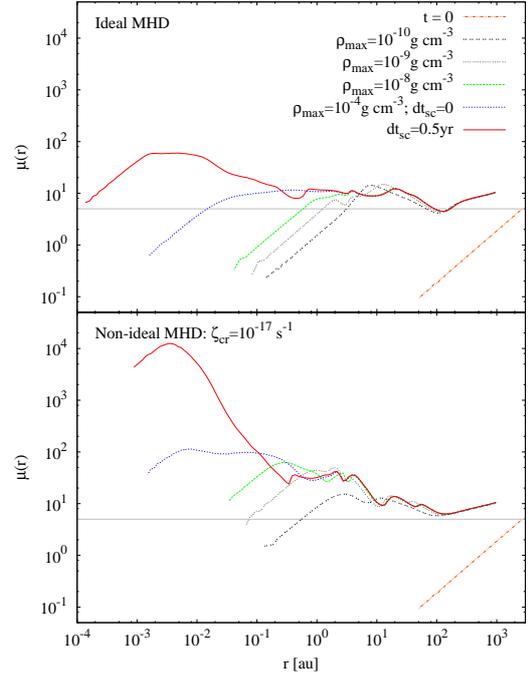}
\caption{The mass-to-flux ratio in units of the critical value as a function of radius at six different epochs for the ideal (top) and non-ideal (bottom) MHD models.  The horizontal grey line represents the initial mass-to-flux ratio, $\mu_0=5$.   At $t=0$, $\mu(r)=\mu_0 = 5$ at $r = R_\text{c}$.  The non-ideal MHD processes diffuse the magnetic field even by \rhoxapprox{-10}, yielding $\mu_\text{non-ideal}(r) > \mu_\text{ideal}(r)$.  There is significant diffusion of the magnetic field for $r \lesssim 10$~au in the non-ideal MHD model.}
\label{fig:mu}
\end{figure}

Fig.~\ref{fig:mu} shows the mass-to-flux ratio (Equation~\ref{eq:masstofluxmu}) for the initial cloud core, at three epochs during the first core phase and two epochs during the stellar core phase.  In the central regions of the core for both the ideal and non-ideal MHD models, the density and magnetic field strength reach a plateau \citep[in agreement with, e.g.,][]{TomidaEtAl2013,WurBatPri2018}; since the mass interior to $r$ necessarily decreases for decreasing $r$, the mass-to-flux ratio $\mu(r)$ also decreases in this region.  Outside of this central region, $\mu(r) > 5$ since the cloud collapses faster along the magnetic field lines than perpendicular to them, increasing the mass as a greater rate than the flux.  During this phase in the non-ideal MHD model, $\mu_\text{non-ideal}(r) > \mu_\text{ideal}(r)$, indicating that the non-ideal processes are diffusing the magnetic field, with significant diffusion for $r \lesssim 10$~au.   The diffusion increases throughout this phase, yielding an increasing maximum $\mu(r)$ as the first core evolves.

During the second collapse phase \citep[$10^{-8}\lesssim \rho/(\text{g cm}^{-3}) \lesssim10^{-4}$;][]{Larson1969}, the central and maximum magnetic field strengths of both models grow as $B \propto \rho_\text{max}^{0.6}$, in agreement with previous studies \citep[e.g.][]{BatTriPri2014,TomOkuMac2015,TsukamotoEtAl2015_oa,WurBatPri2018,VaytetEtAl2018}.  During this growth, the maximum magnetic field strength and maximum density are coincident for the ideal MHD model, but not the non-ideal MHD model.

\begin{figure}
\centering
\includegraphics[width=0.8\columnwidth]{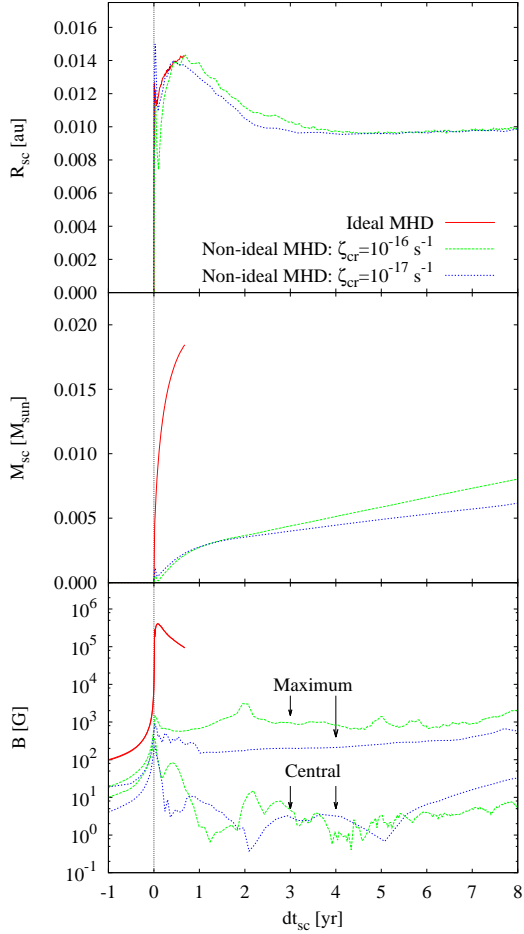}
\caption{The time evolution of the radius and mass of the stellar core after the formation of the stellar core (at \dtsczero \ when \rhoxeq{-4}); the bottom panel shows the maximum magnetic field strength (which resides outside the stellar core for the non-ideal MHD model), and the central strength (coincident with \rhox).  The three models are ideal MHD, our fiducial non-ideal MHD with \zetaeq{-17}, and a non-ideal MHD model with \zetaeq{-16}.  The stellar core radii are approximately independent of the non-ideal processes, however, their mass growth rate is dependent, with the growth rate slowing considerably for the non-ideal MHD models.  The magnetic field strength in the stellar core of the ideal MHD model is higher than those found in young stars, and for the entire simulation, $B_\text{max} = B_\text{cen}$.  The non-ideal MHD models have central magnetic field strengths below \sm30~G over the first \dtsc{8}, suggesting that the magnetic fields cannot be fossil in origin once non-ideal MHD processes are self-consistently modelled.}
\label{fig:Vtime:sc}
\end{figure}

We define the formation of the stellar core, \dtsczero, to be when \rhoxeq{-4}.  We define the radius of the core using the gas with \rhoge{-4}.  Due to computational limitations, the ideal MHD model is evolved until \dtsc{0.7}, while the non-ideal MHD model is evolved until 11.4~yr after the formation of the stellar core.  The top two panels of Fig.~\ref{fig:Vtime:sc} show the evolution of the radius and mass of the stellar core; the bottom panel shows the evolution of the central and maximum magnetic field strengths.  Although both the ideal and non-ideal MHD models have similar radii (at least for \dtscls{0.7}), the mass contained within the stellar core differs, with the ideal model reaching a mass of 0.018\Msun \ at 0.7~yr, whereas the non-ideal MHD model reaches a mass of only 0.0083\Msun \ by 11.4~yr.  

\begin{figure*} 
\centering
\includegraphics[width=0.31\textwidth,trim={0           0 0 0}]{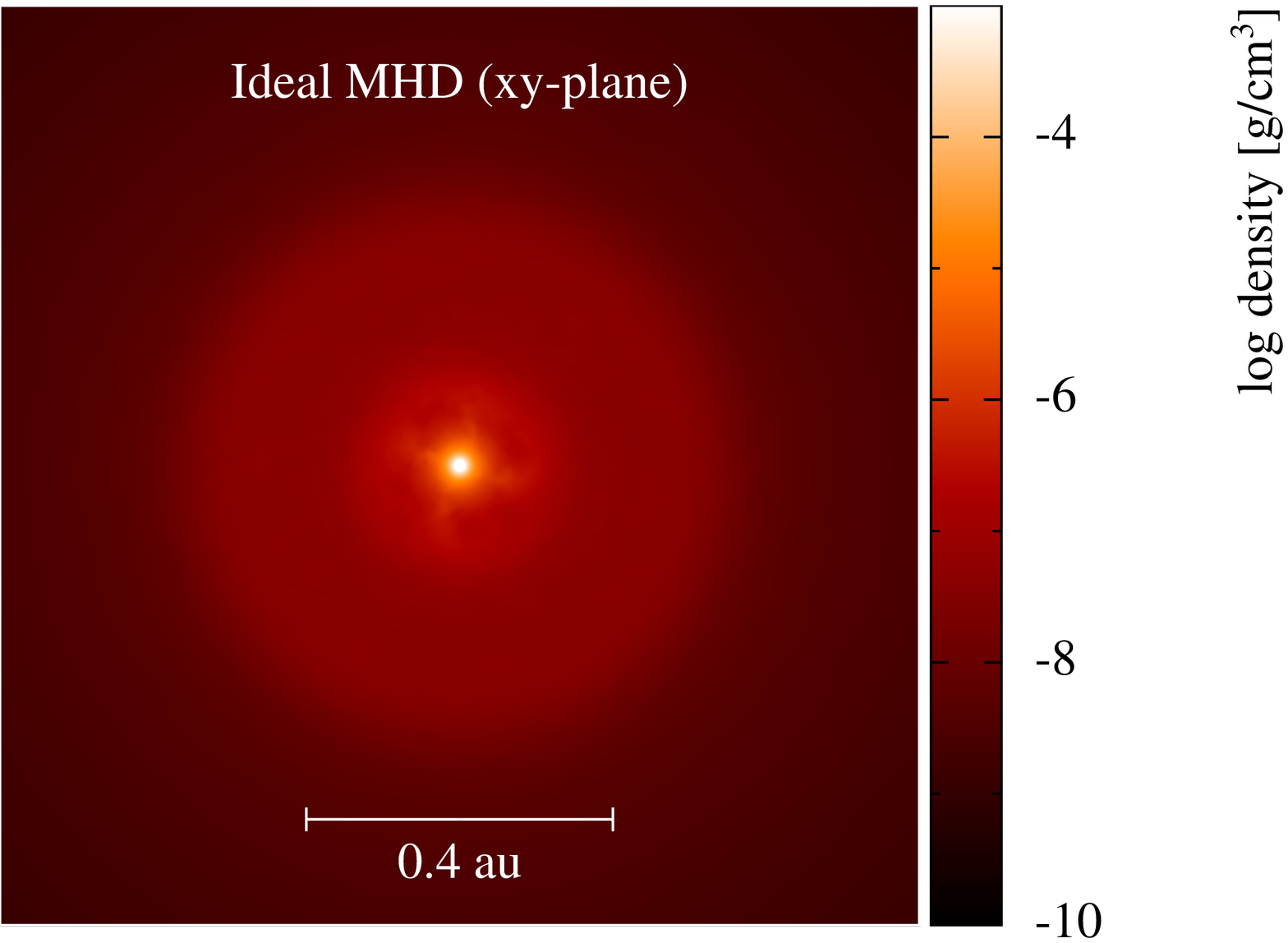}
\includegraphics[width=0.22\textwidth,trim={5.85cm 0 0 0}]{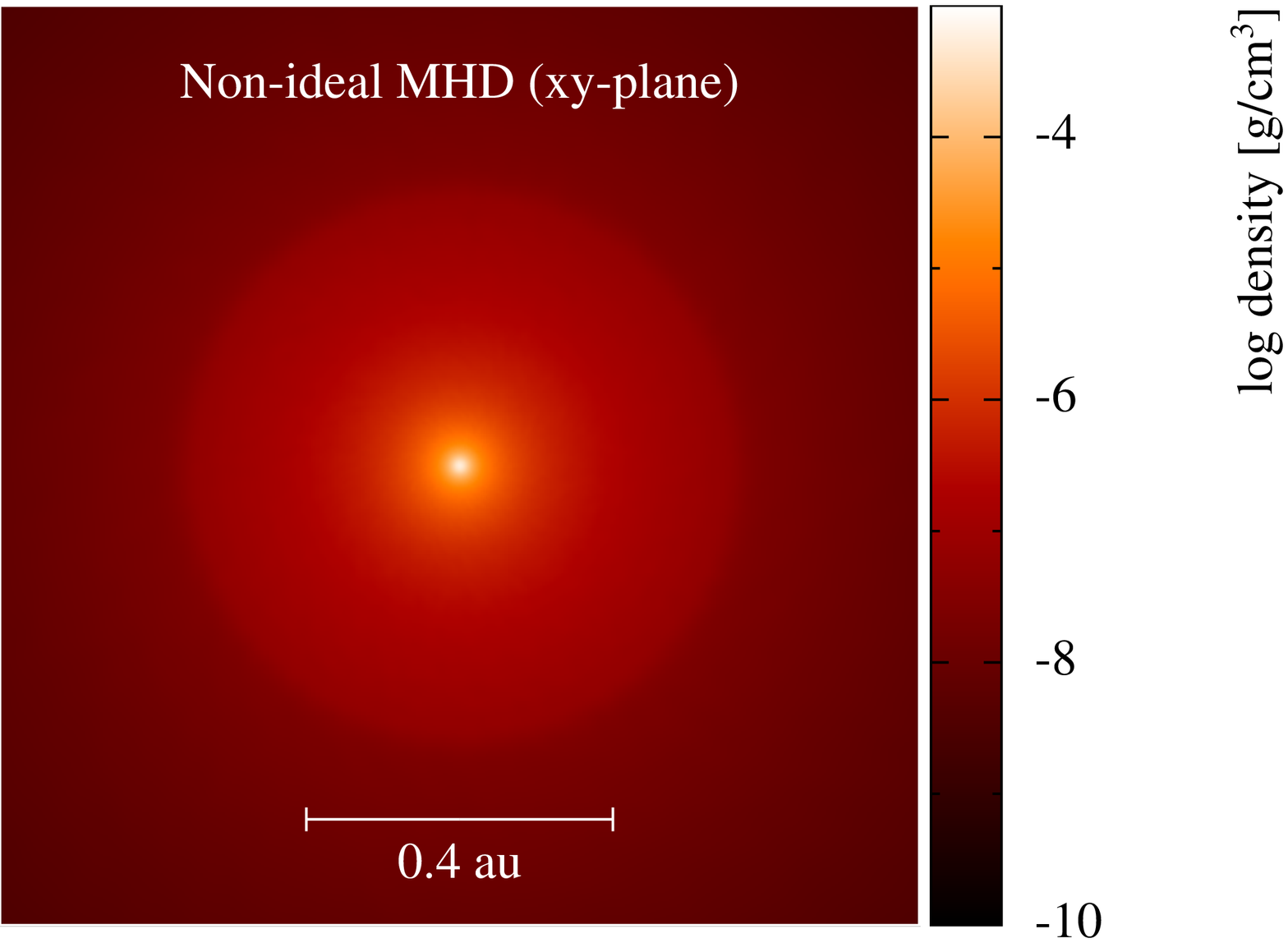}
\includegraphics[width=0.22\textwidth,trim={5.85cm 0 0 0}]{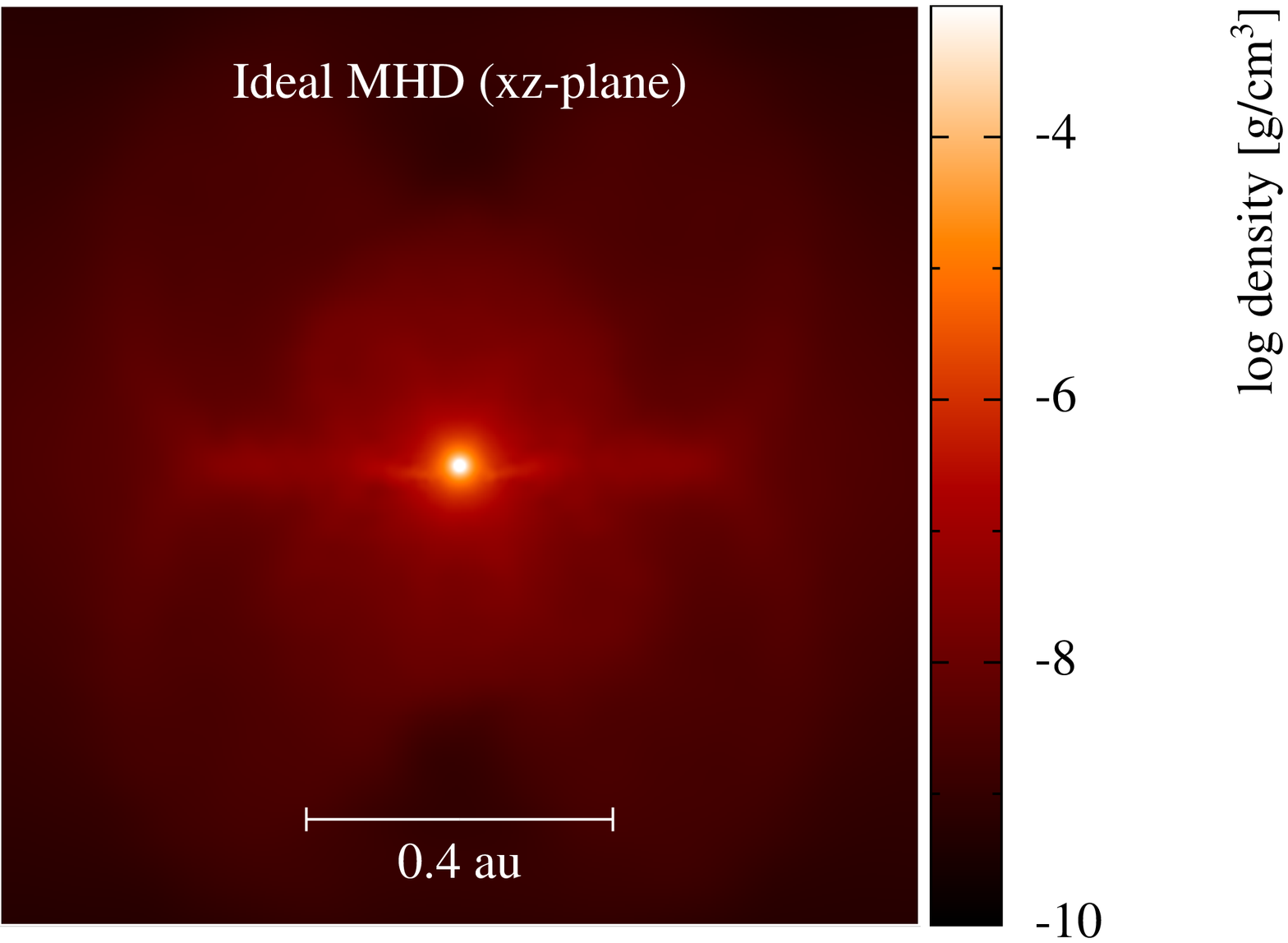}
\includegraphics[width=0.22\textwidth,trim={5.85cm 0 0 0}]{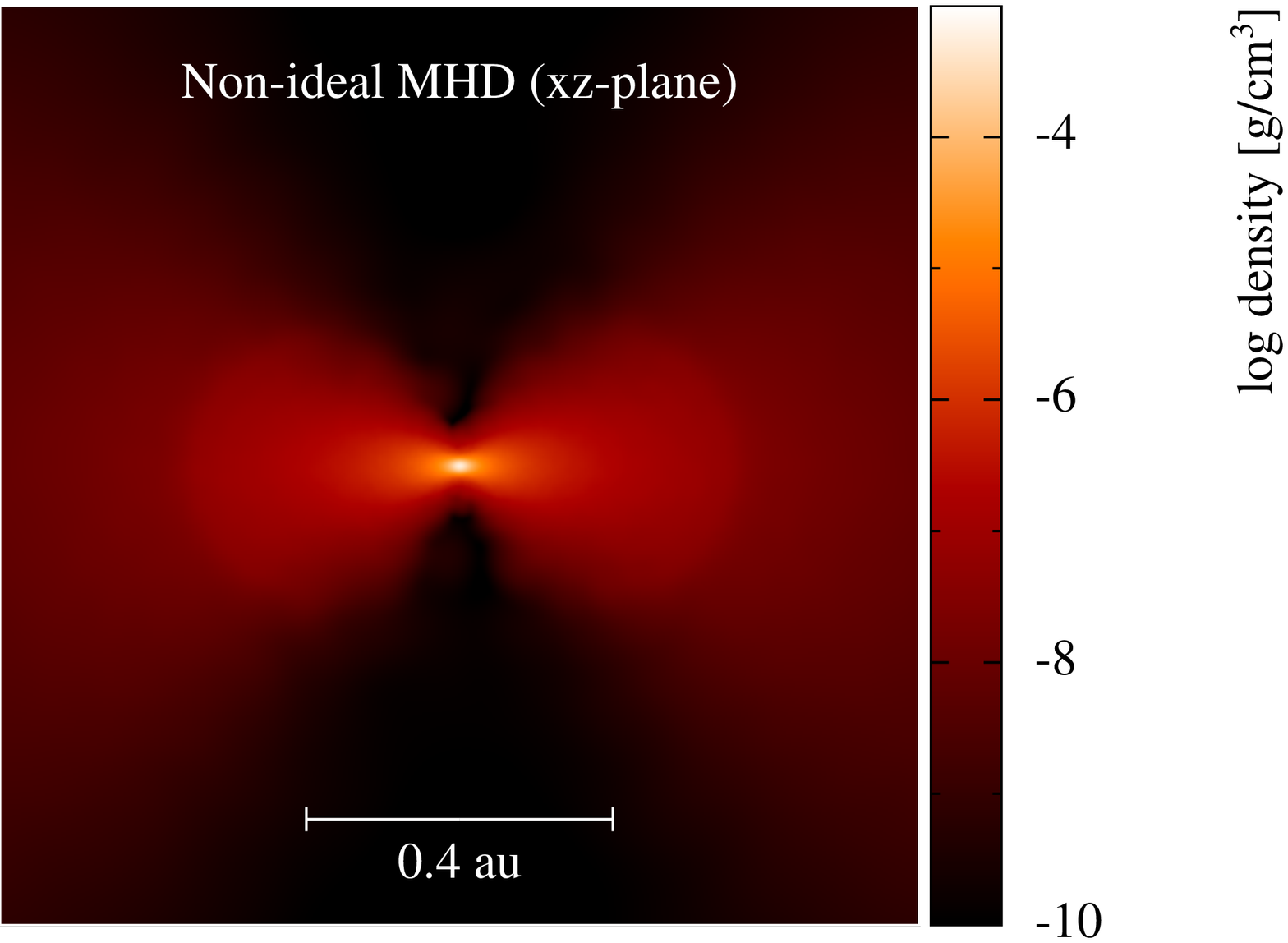}
\includegraphics[width=0.31\textwidth,trim={0           0 0 0}]{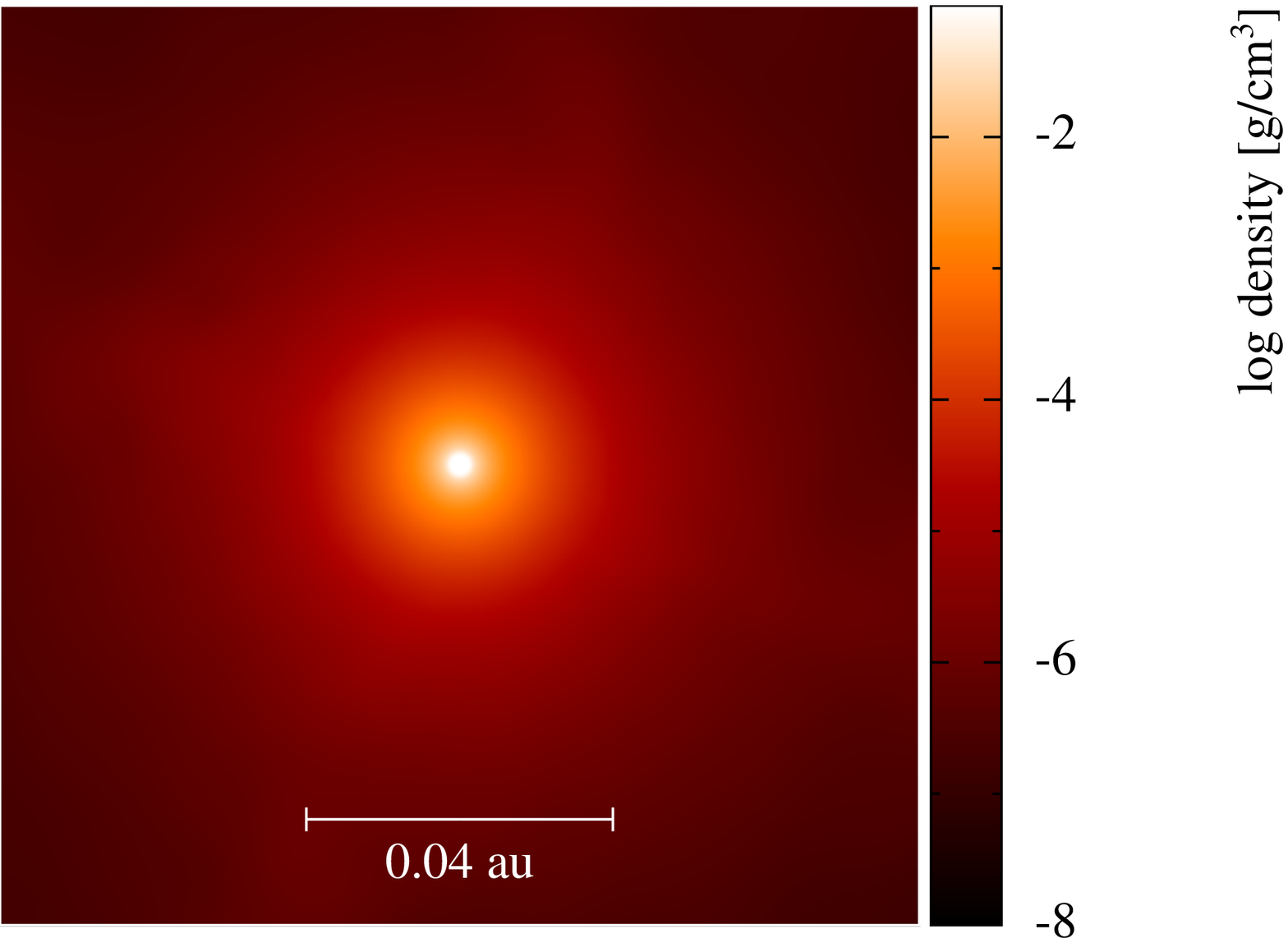}
\includegraphics[width=0.22\textwidth,trim={5.85cm 0 0 0}]{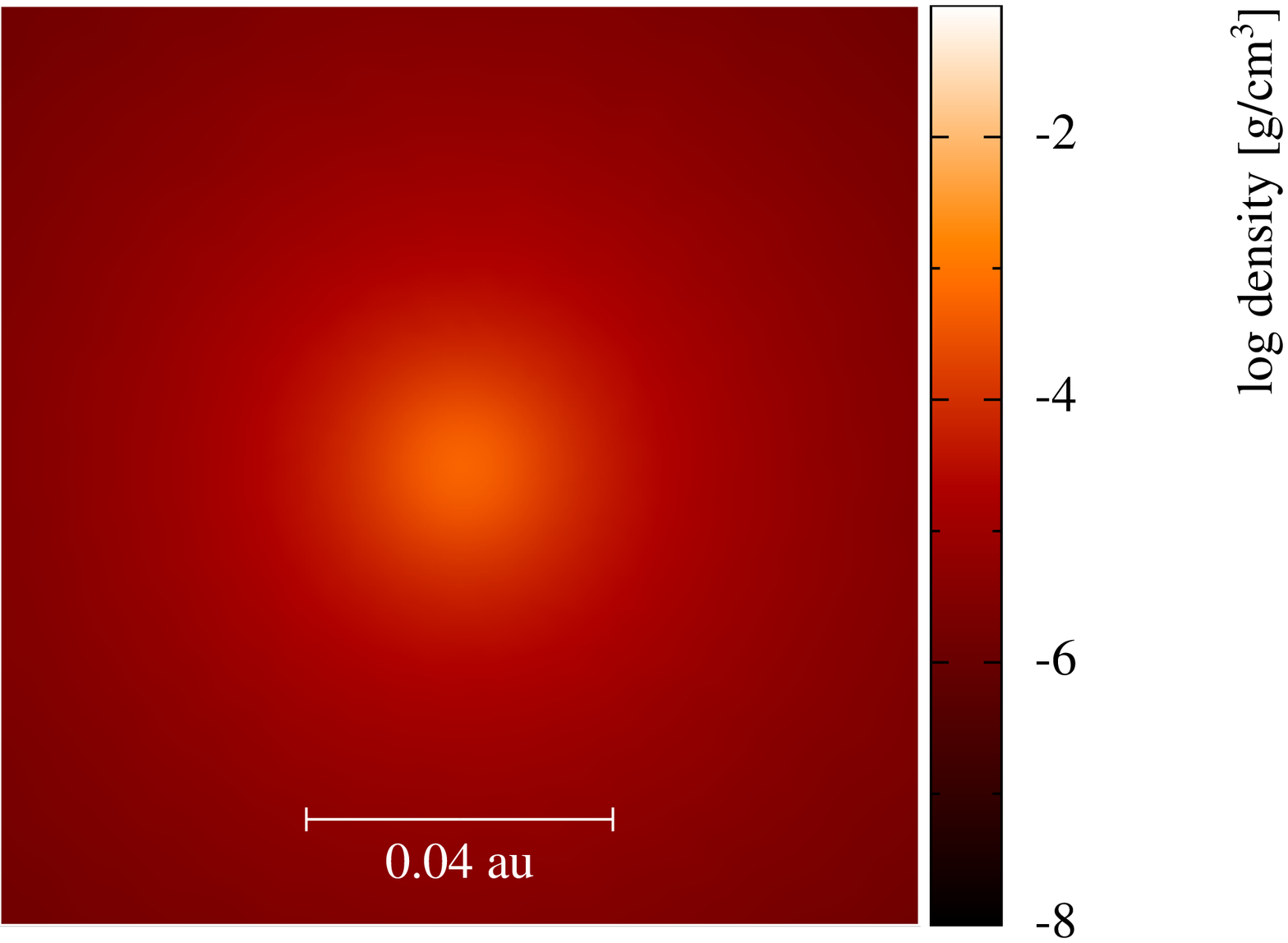}
\includegraphics[width=0.22\textwidth,trim={5.85cm 0 0 0}]{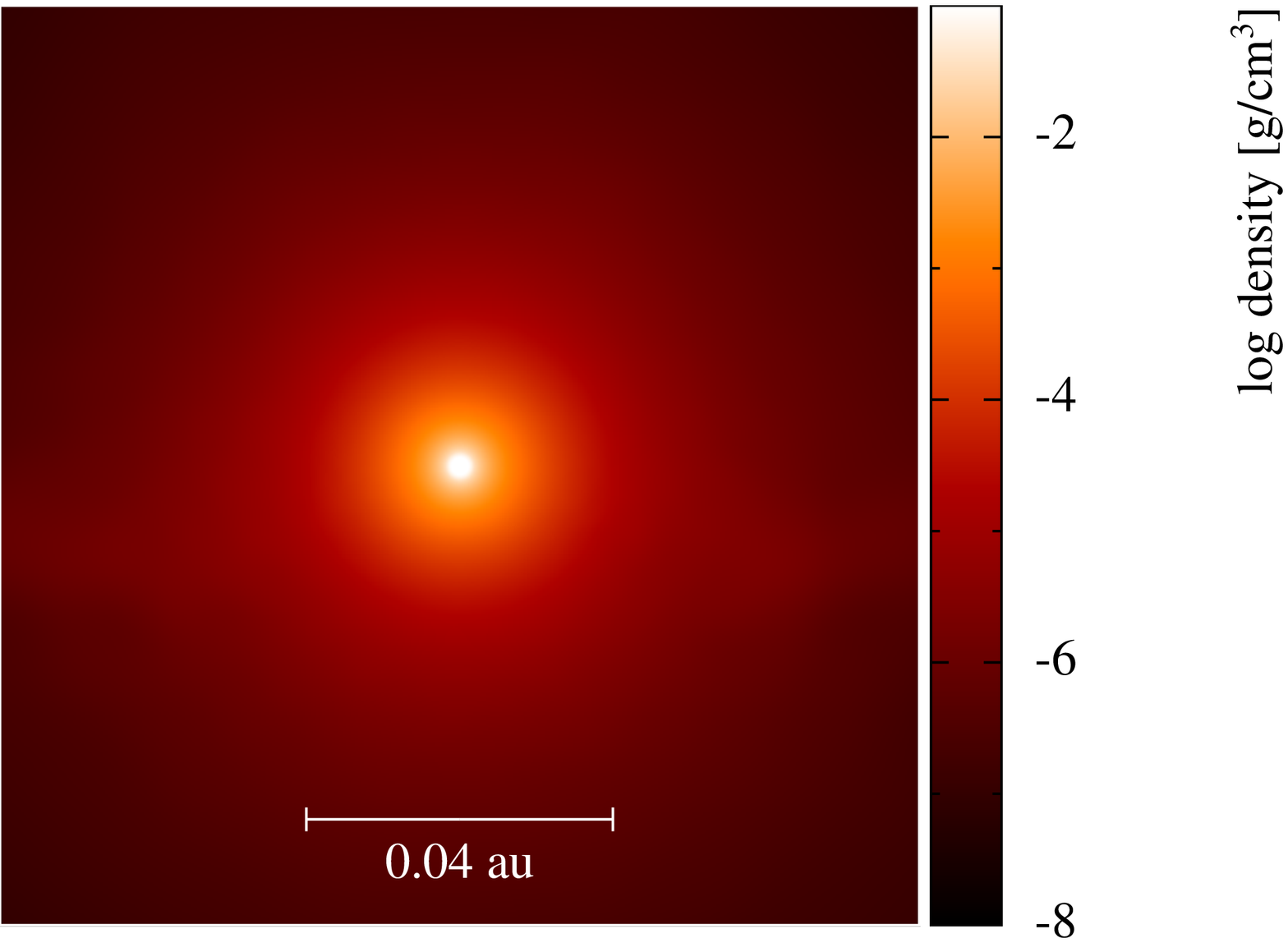}
\includegraphics[width=0.22\textwidth,trim={5.85cm 0 0 0}]{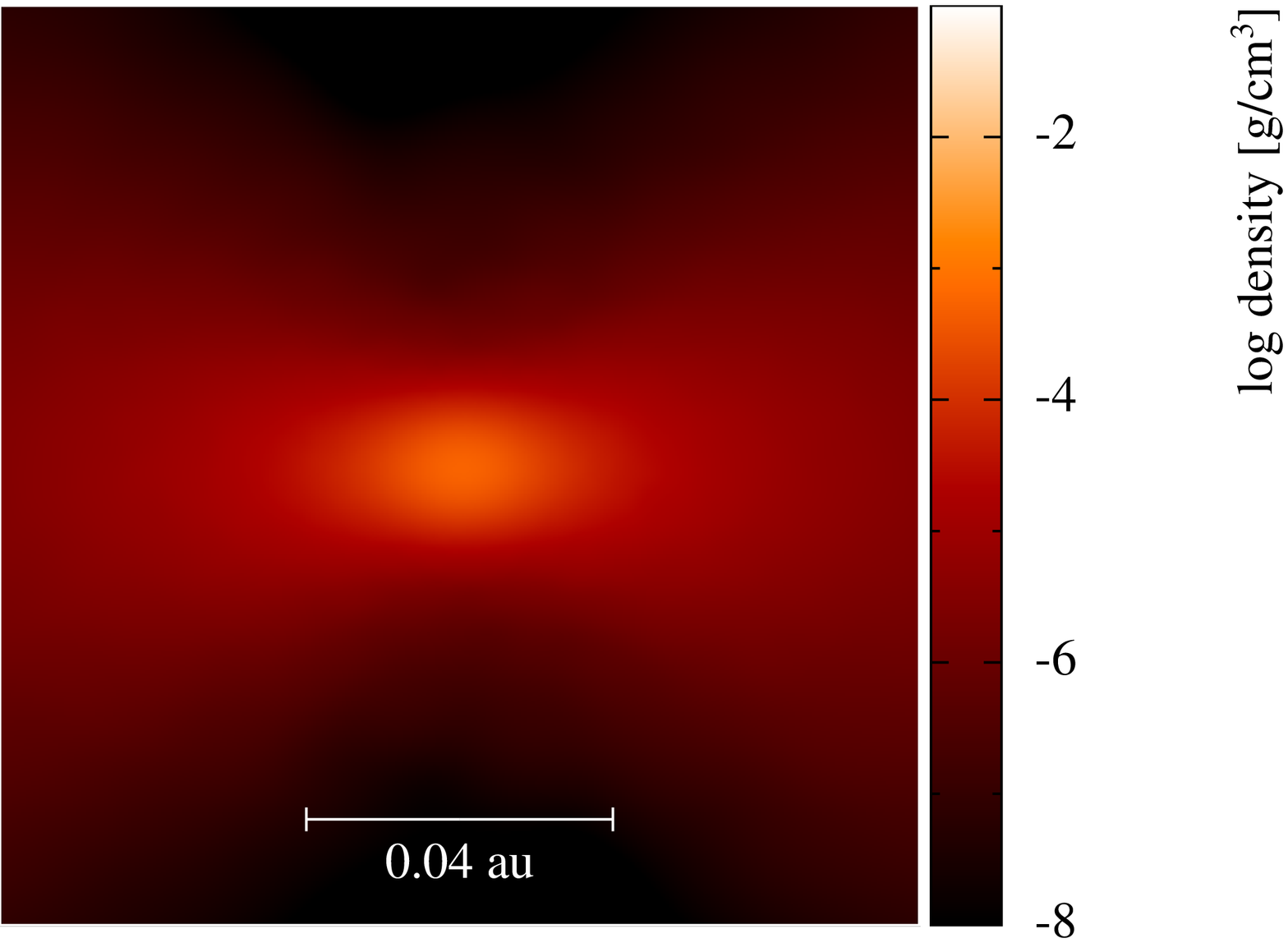}
\includegraphics[width=0.31\textwidth,trim={0           0 0 0}]{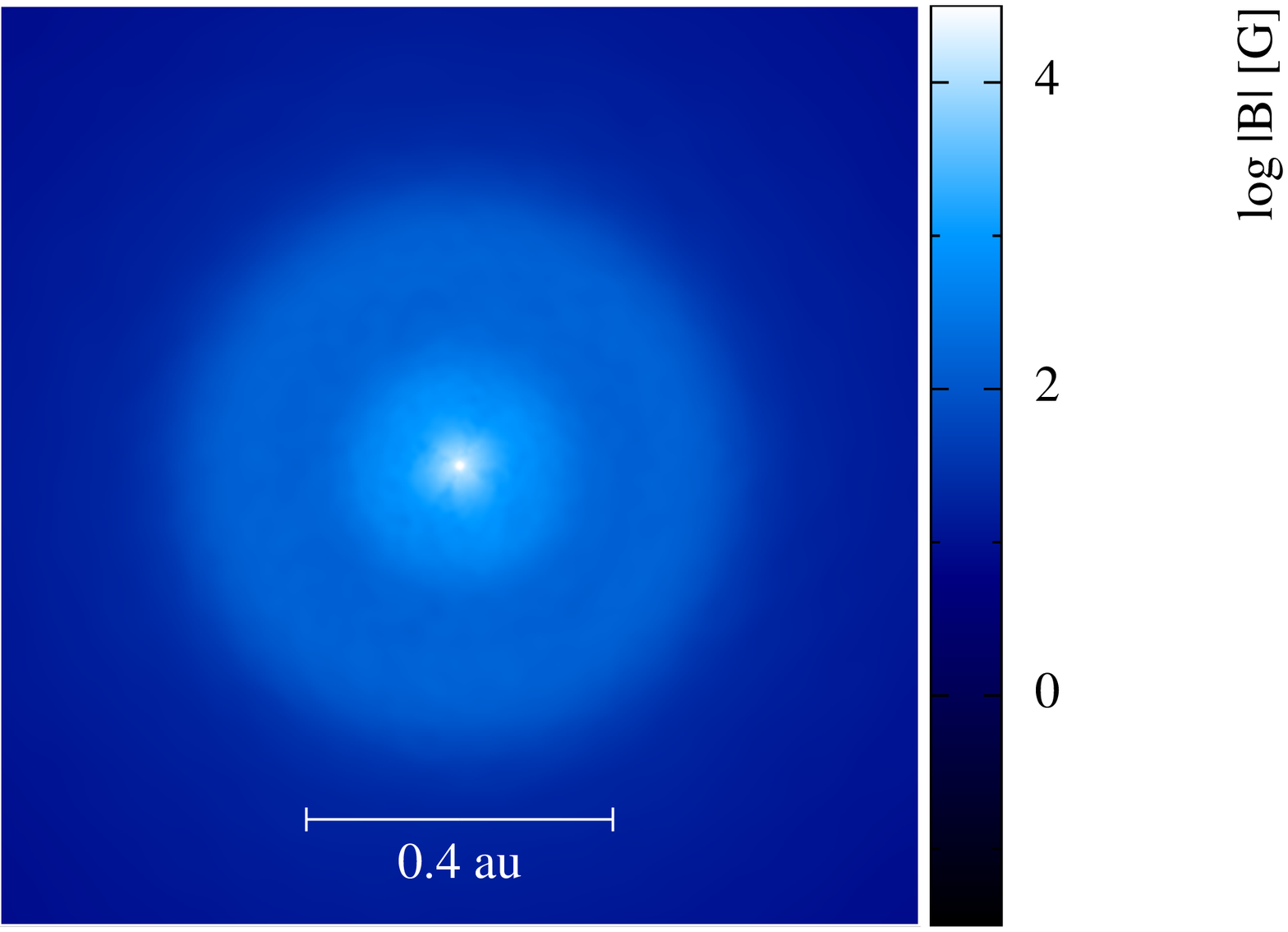}
\includegraphics[width=0.22\textwidth,trim={5.85cm 0 0 0}]{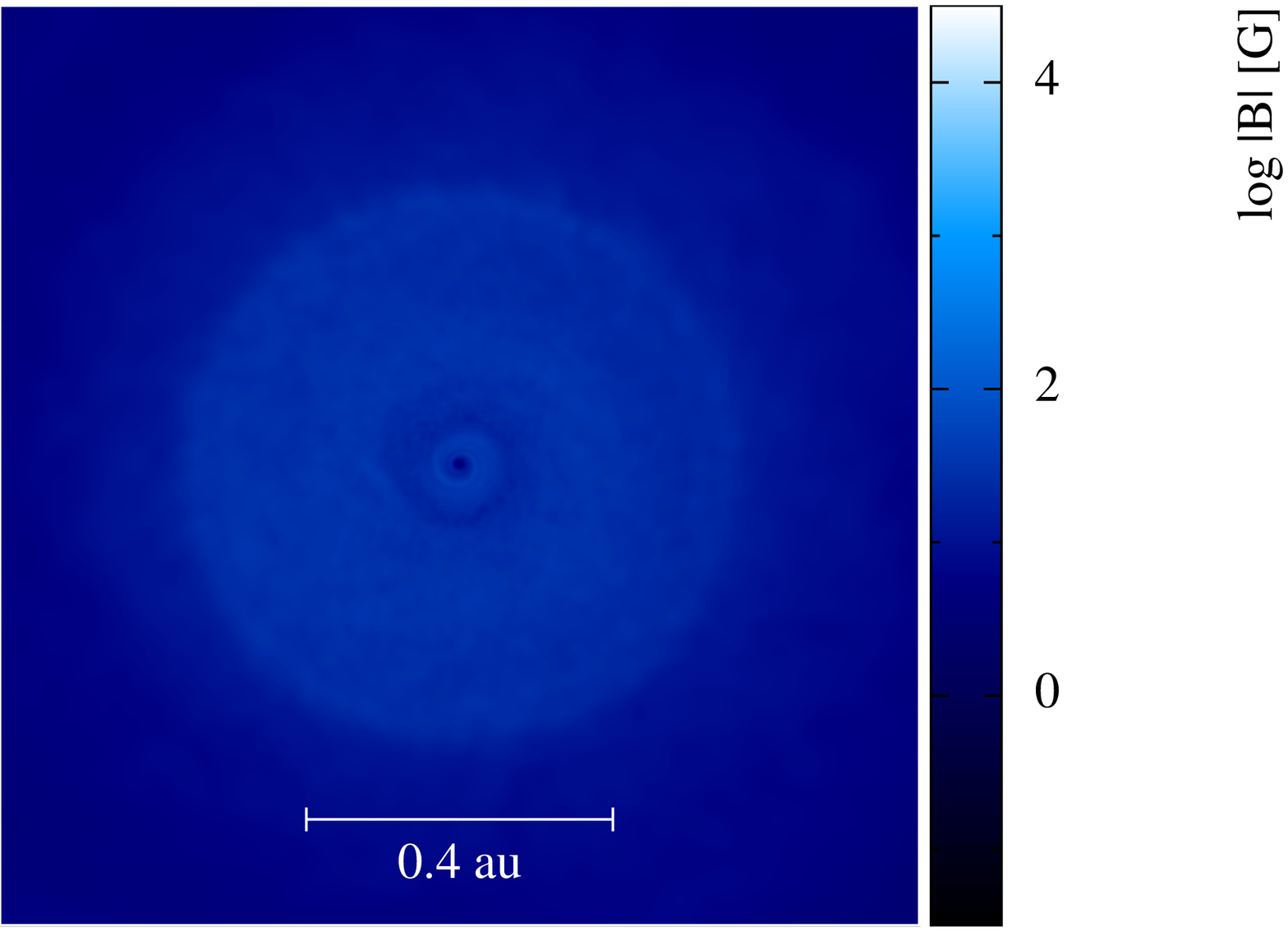}
\includegraphics[width=0.22\textwidth,trim={5.85cm 0 0 0}]{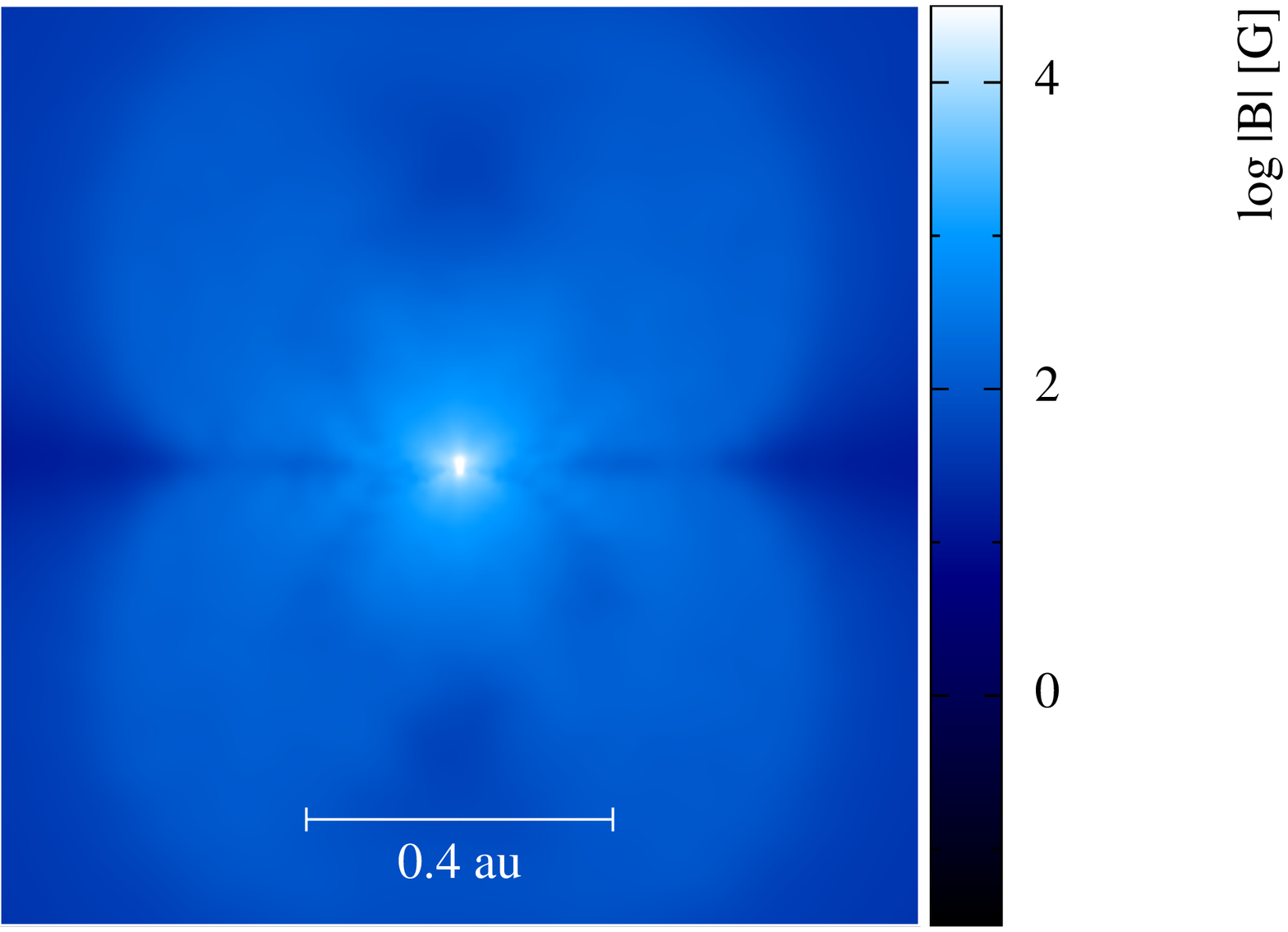}
\includegraphics[width=0.22\textwidth,trim={5.85cm 0 0 0}]{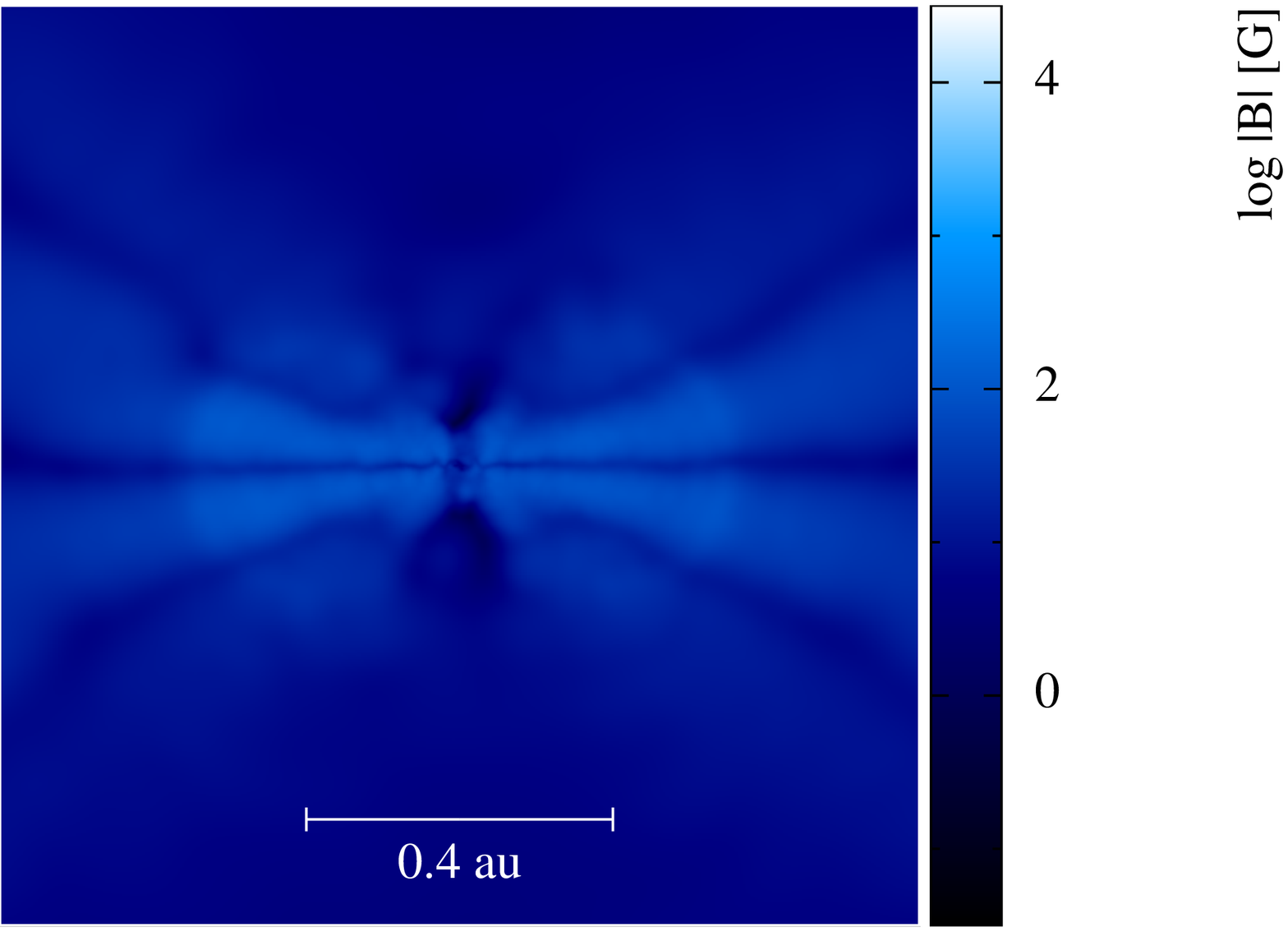}
\includegraphics[width=0.31\textwidth,trim={0           0 0 0}]{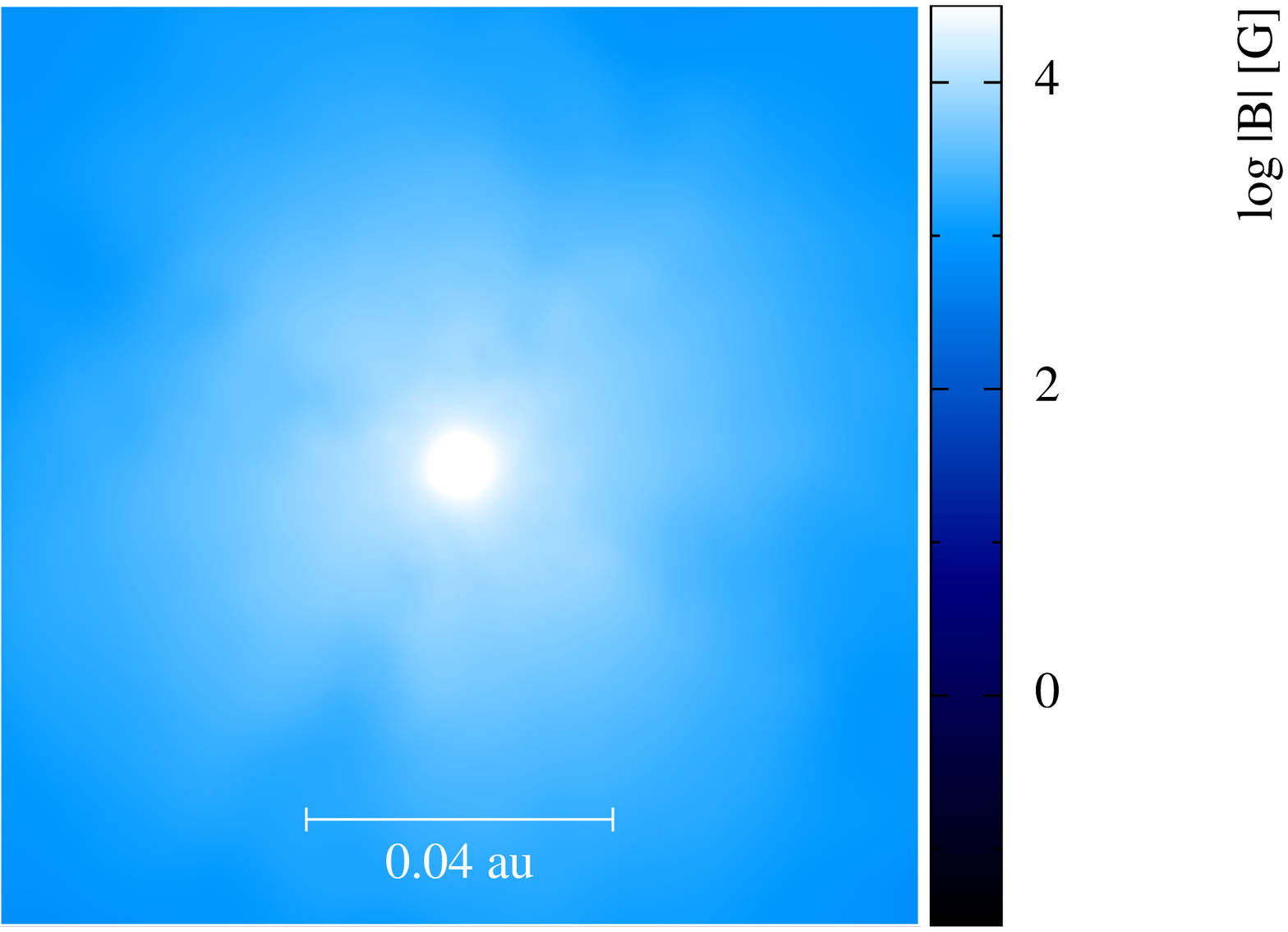}
\includegraphics[width=0.22\textwidth,trim={5.85cm 0 0 0}]{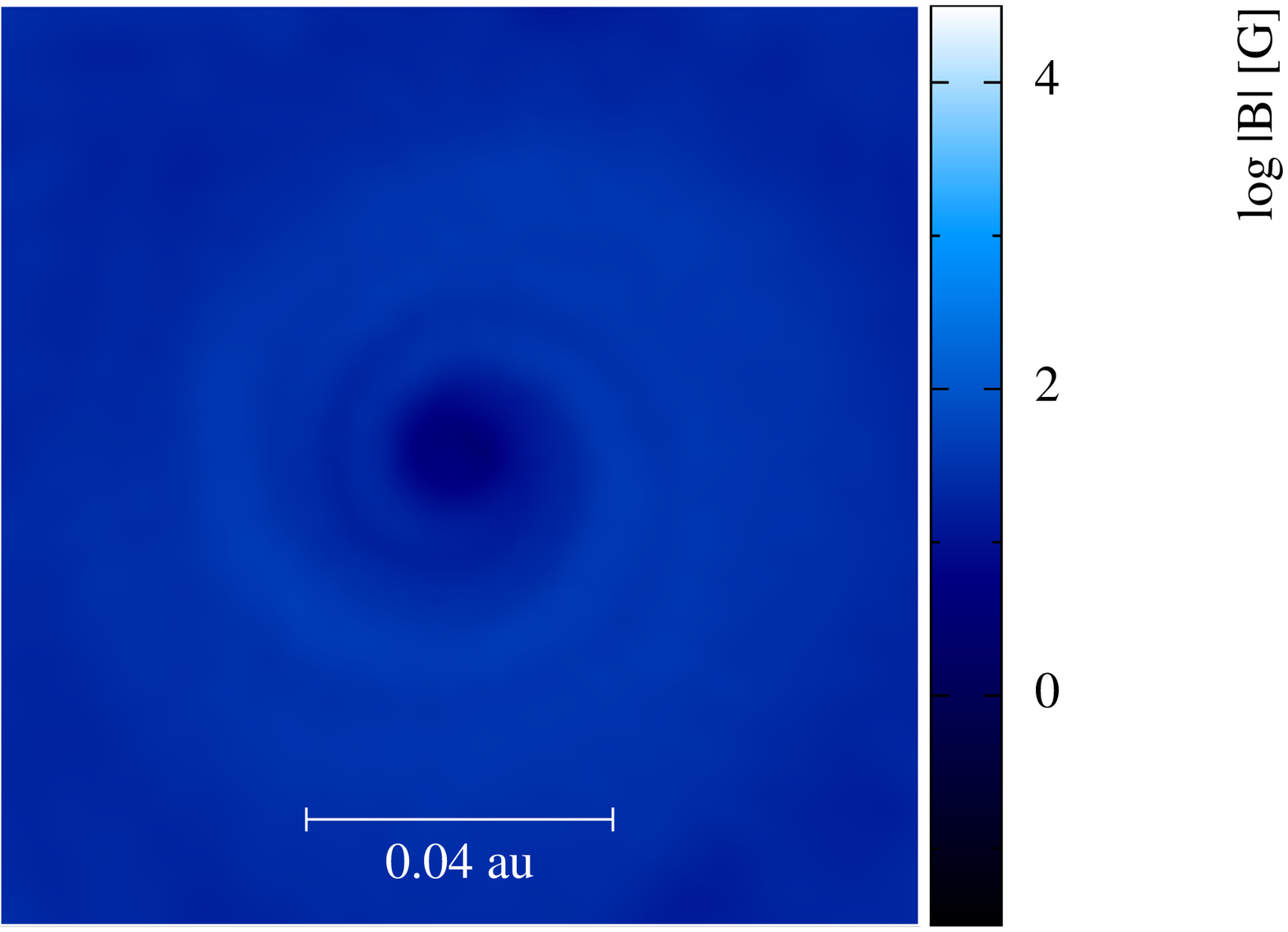}
\includegraphics[width=0.22\textwidth,trim={5.85cm 0 0 0}]{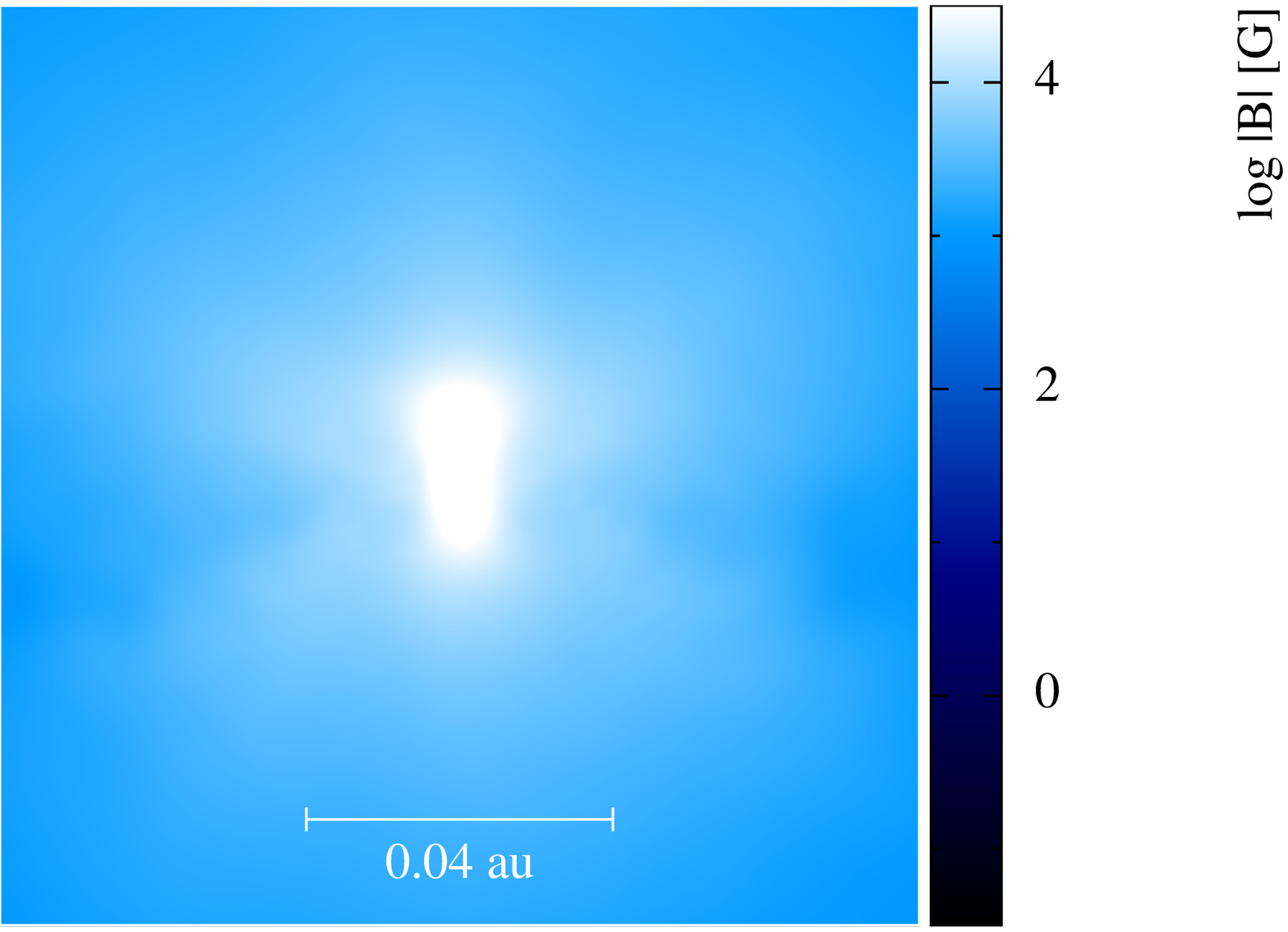}
\includegraphics[width=0.22\textwidth,trim={5.85cm 0 0 0}]{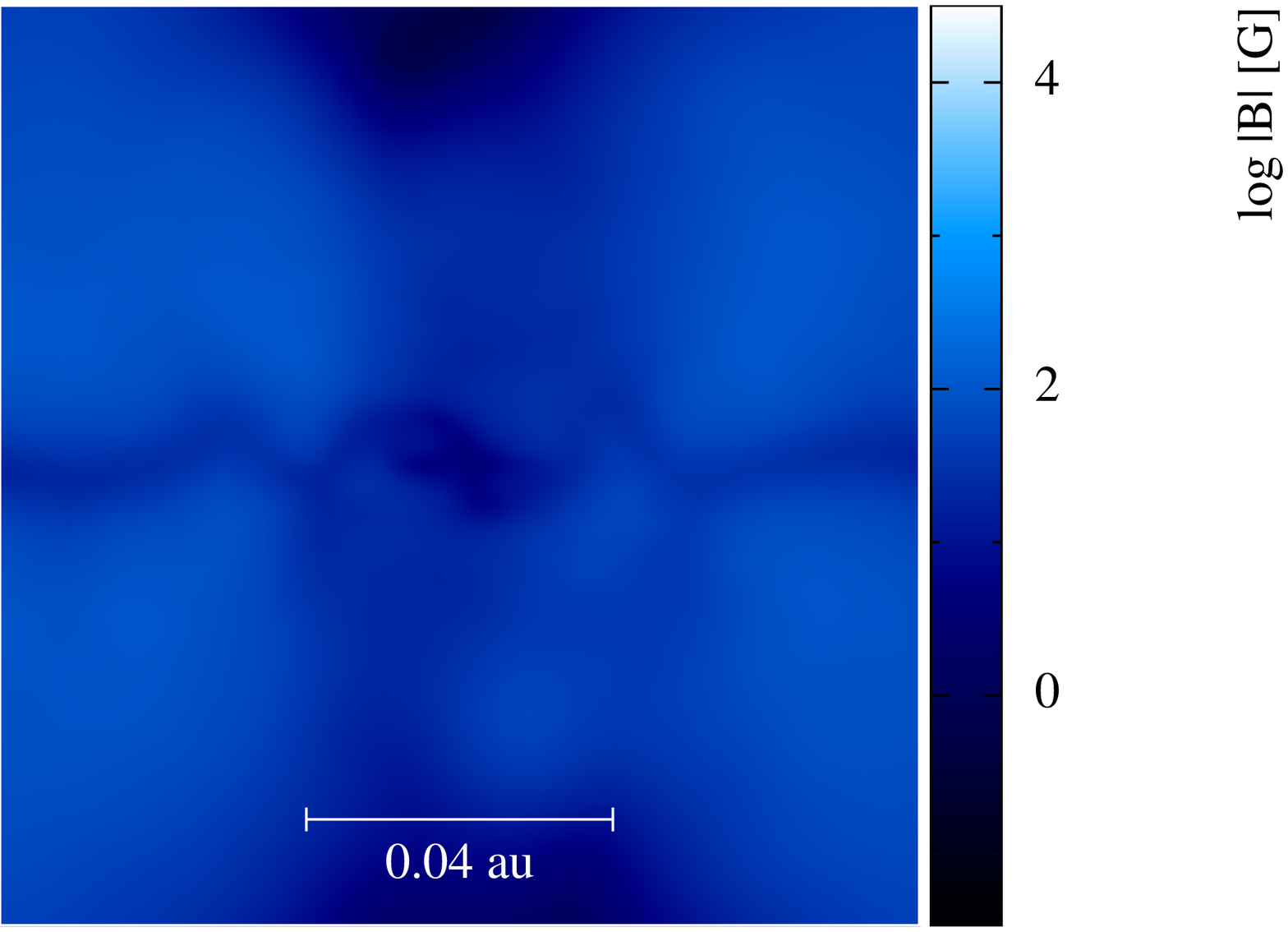}
\caption{Stellar core structure at \dtsc{0.5}:  Density (top two rows) and magnetic field strength (bottom two row) slices through the stellar core perpendicular (first two columns) and parallel (final two columns) to the rotation axis at two different spatial scales, for the ideal and non-ideal (\zetaeq{-17}) MHD models.  By this time, the ideal MHD model has reached a maximum density in excess of $\rho_\text{max} \approx 0.1$ g cm$^{-3}$, and the maximum magnetic field is in the centre of the stellar core.  The stellar core in the non-ideal MHD model grows much more slowly, and the maximum magnetic field strength in this model lies outside the stellar core.  The stellar core magnetic field strength is \sm$3\times10^4$ times lower in the non-ideal MHD model compared to the ideal MHD model.}
\label{fig:rhobfield}
\end{figure*} 

A dense stellar core with a strong magnetic field forms in the ideal MHD model.  The maximum magnetic field strength continues to increases until  \rhoxapprox{-1}, after which $B_\text{max}$ to decreases and $\mu(r)$ increases due to artificial resistivity; see the solid red line in the top panel of Fig.~\ref{fig:mu} above and Section~\ref{sec:numresis} below.  The maximum magnetic field strength reached in the stellar core occurs several days after its formation and is $B_\text{max}\sim4\times10^5$~G.  This magnetic field strength is much stronger than the $B\sim10^3$~G field found in young stars \citep[e.g.][]{JohValHatKan1999,ValJoh2004,SymHarKurNay2005,YanJohVal2007}.  During the entire evolution of the ideal MHD model, the maximum magnetic field strength is in the centre of the core.

Fig.~\ref{fig:rhobfield} shows cross sections of the density and magnetic field strength in the stellar cores at \dtsc{0.5}, both parallel and perpendicular to the rotation axis and at two difference spatial scales.  At this time, the ideal MHD model has a spherical stellar core, with a magnetic field strength that decreases with distance from the centre of the core.  The elongated region of strong magnetic field strength corresponds to the stellar core outflow.

At the formation of the stellar core, the central magnetic field strength in the non-ideal MHD model is \sm60 times lower than in the ideal MHD model  (see blue lines in Fig.~\ref{fig:mu} and bottom panel of Fig.~\ref{fig:Vtime:sc}).  This difference is a direct result of the different magnetic field structure produced during the first hydrostatic core phase.  

After the formation of the stellar core, the reduced magnetic braking and resulting greater angular momentum leads to the formation of a disc (Fig.~\ref{fig:rhobfield}) in the non-ideal MHD model.  The maximum magnetic field in this disc is located $\gtrsim 0.02$~au from the centre (the stellar core has a radius of $\sim$0.01~au  or 2~R$_\odot$) and reaches a maximum of \sm900~G several days after the formation of the stellar core.  Thus, the maximum magnetic field strength continues to be in the disc rather than the stellar core itself, and the maximum magnetic field never again becomes coincident with the centre of the stellar core.  The maximum magnetic field strength remains below $\sim$900~G, which is $\sim$100--500 times lower that of the ideal MHD model.  Thus, even when considering the maximum magnetic field strength, this model rules out the formation of strong fossil fields.

At the birth of the stellar core, the central magnetic field in the non-ideal MHD model is  $B_\text{cen}$ \sm170~G and reaches \sm240~G a few days later.  It then decreases to \sm4~G after \dtsc{0.5}.  As the stellar core evolves, the central magnetic field strength fluctuates, but remains below 10~G within the first 6~yr after the formation of the stellar core.  The strength rises to \sm30~G after \dtscapprox{8}, but does not surpass this value for the duration of the simulation (ending at \dtscapprox{11}).  In the stellar core, the magnetic Reynolds number is $R_\text{m,art} \sim 100$ for artificial resistivity and $R_\text{m,phys} \sim 10^8$ for physical resistivity. Since $R_\text{m,art} \ll R_\text{m,phys}$, any diffusion that is occurring in the stellar core is due to the artificial resistivity (see Section~\ref{sec:numresis}).

The final central magnetic field strength is much less than the kG magnetic field strengths observed in young stars, implying that the magnetic fields in low-mass stars are generated in a subsequent dynamo process rather than being fossil in origin.  Even if the weak magnetic field is a result of artificial resistivity, the central magnetic field of $B_\text{cen}\sim170-240$~G implanted at the birth of the stellar core is still below the observed kG magnetic field strengths, thus providing evidence against a fossil field origin and favouring the dynamo process.

%-----
\begin{figure*}
\centering
\includegraphics[width=0.8\textwidth]{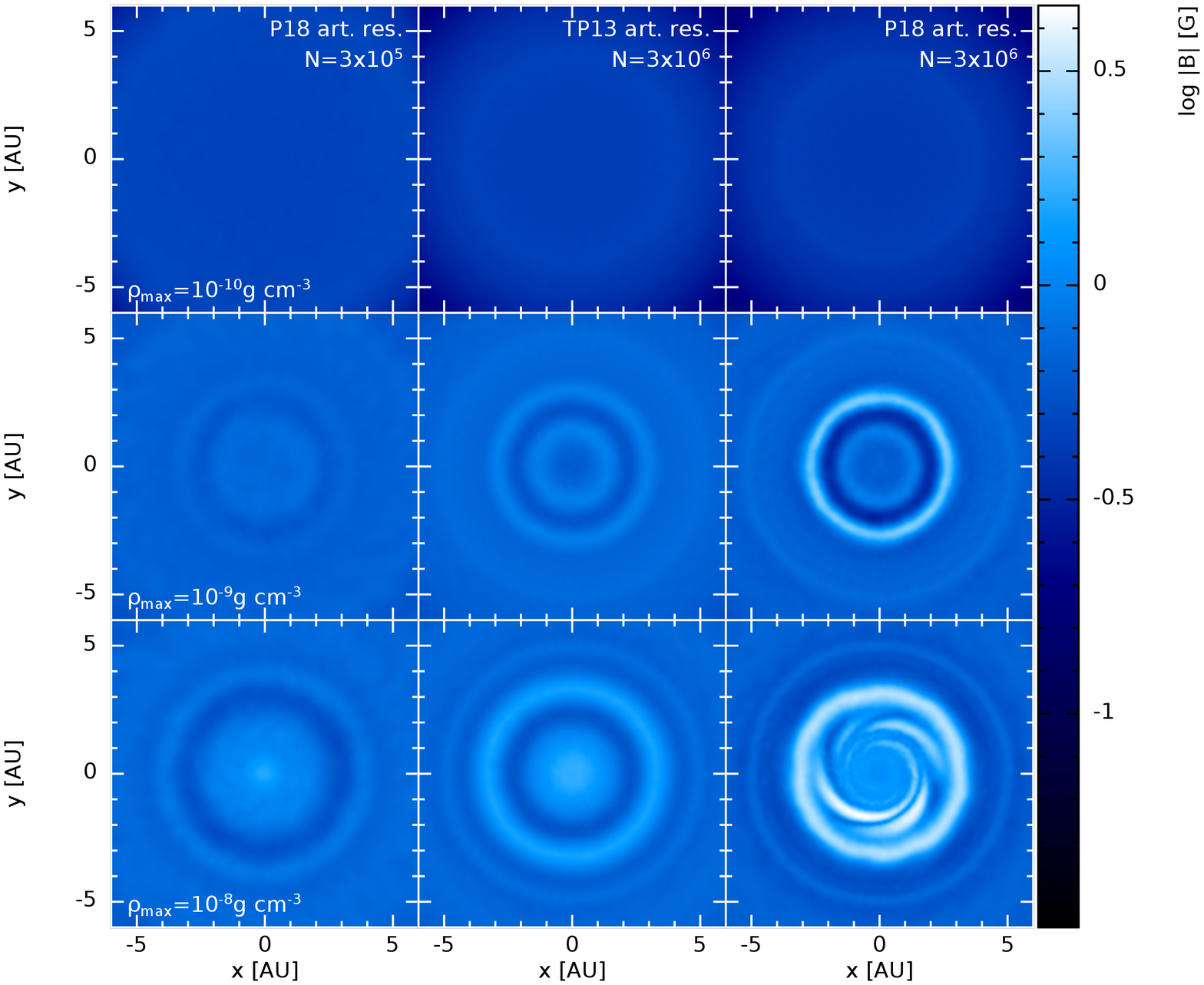}
\caption{Testing the dependence of the magnetic field structure in the first core phase on resolution and artificial resistivity for the non-ideal MHD model with \zetaeq{-17}.  For our resolution study, the model in the left-hand column are initialised with $3\times10^5$ particles in the initial sphere, and the remaining two columns are initialised with $3\times10^6$ particles.  To test artificial resistivity, we use the algorithm from \citet{TriPri2013} (centre column) and \citet{Phantom2017} (left- and right-hand columns).  The panels are slices of magnetic field strength through the first core, perpendicular to the rotation axis at $\rho_\text{max} \approx 10^{-10}$ (top), $10^{-9}$ (middle) and $10^{-8}$~g cm$^{-3}$ (bottom).  The magnetic field piles up in a torus at the edge of the first core in all three models, resulting in a weak central magnetic field strength.  However, the strong artificial resistivity in the $3\times10^5$ model, and to a lesser extent in the TP13 model, diffuses the magnetic torus, resulting in a more uniform structure and a slightly stronger central magnetic field strength.}
\label{fig:slice:B:fhc}
\end{figure*}

%--------------------------------------------------------------------------------
\section{Discussion}
\label{sec:dis}
%----------
\subsection{Effect of higher cosmic ray ionization rates}
As the comic ray ionization rate is increased, the gas should become more ionized, resulting in a stronger magnetic field.  Thus, to verify our conclusion, we model an additional non-ideal MHD model using the higher cosmic ray ionization rate of \zetaeq{-16}, which is 10 times higher than the canonical value.

Similar to the model with \zetaeq{-17}, the maximum and central magnetic field strengths are no longer coincident during the first core phase, and, as expected, remain higher than their counterpart strengths in the model with \zetaeq{-17}.  During the stellar core evolution phase, the central magnetic field strengths decrease in both models and remain below $\sim$30~G (bottom panel of Fig.~\ref{fig:Vtime:sc}).  Thus, even by increasing the cosmic ray ionization rate to 10 times the canonical value, the stellar core magnetic field strength is still several orders of magnitude below that required for fossil fields.

%----------
\subsection{Effect of resolution and artificial resistivity}
\label{sec:numresis}
\begin{figure*}
\centering
\includegraphics[width=0.8\textwidth]{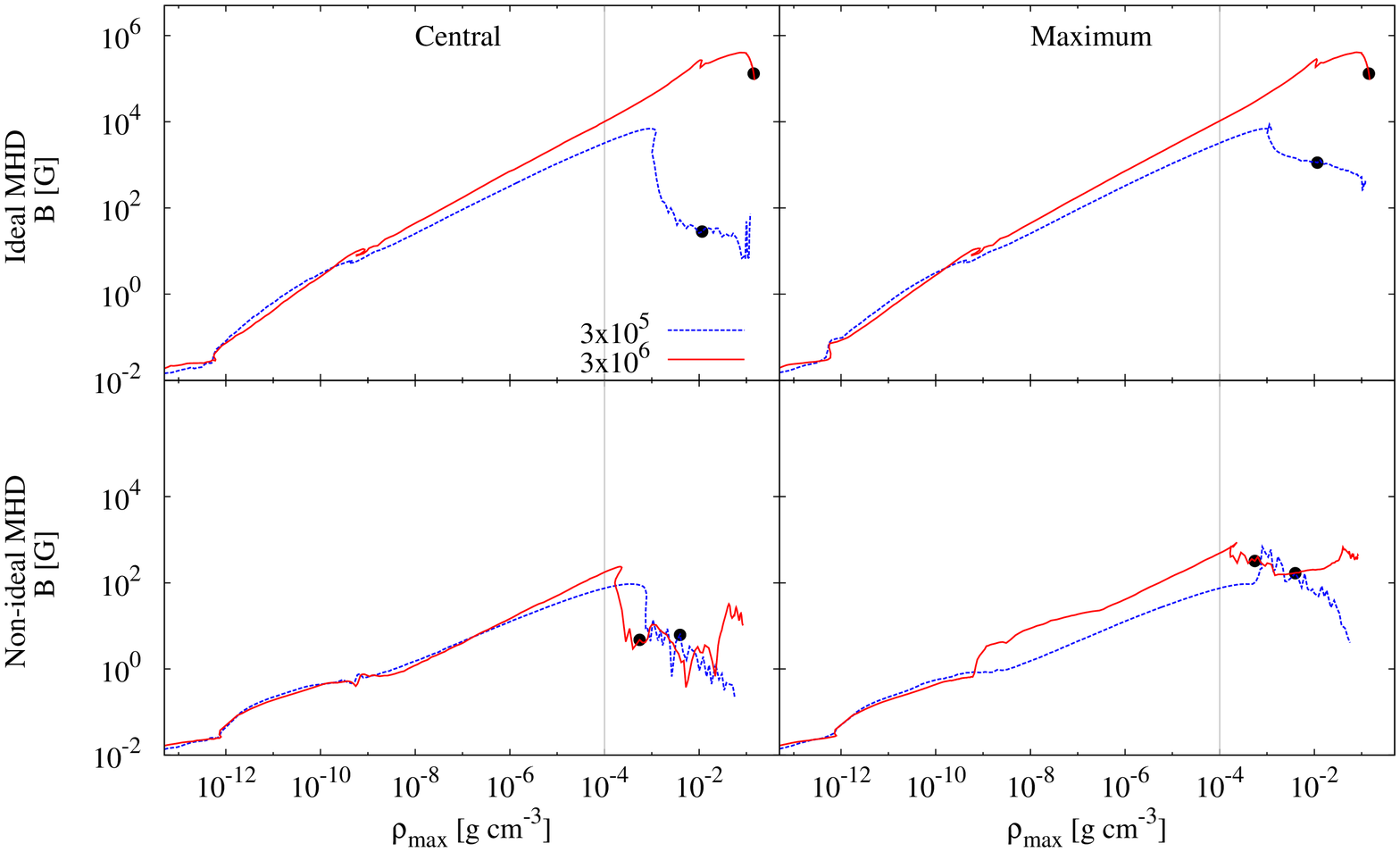}
\caption{The time evolution of the central (left) and maximum (right) magnetic field strengths at two different resolutions.  The non-ideal MHD models use \zetaeq{-17}.  The vertical grey line represents the formation density of the stellar core, and the black circles are placed at \dtsc{0.5}.  The magnetic field evolution is dependent on resolution for ideal MHD, whereas the central magnetic field strength in the non-ideal model is approximately independent of resolution.}
\label{fig:BVrho:res}
\end{figure*}
\begin{figure*}
\centering
\includegraphics[width=0.8\textwidth]{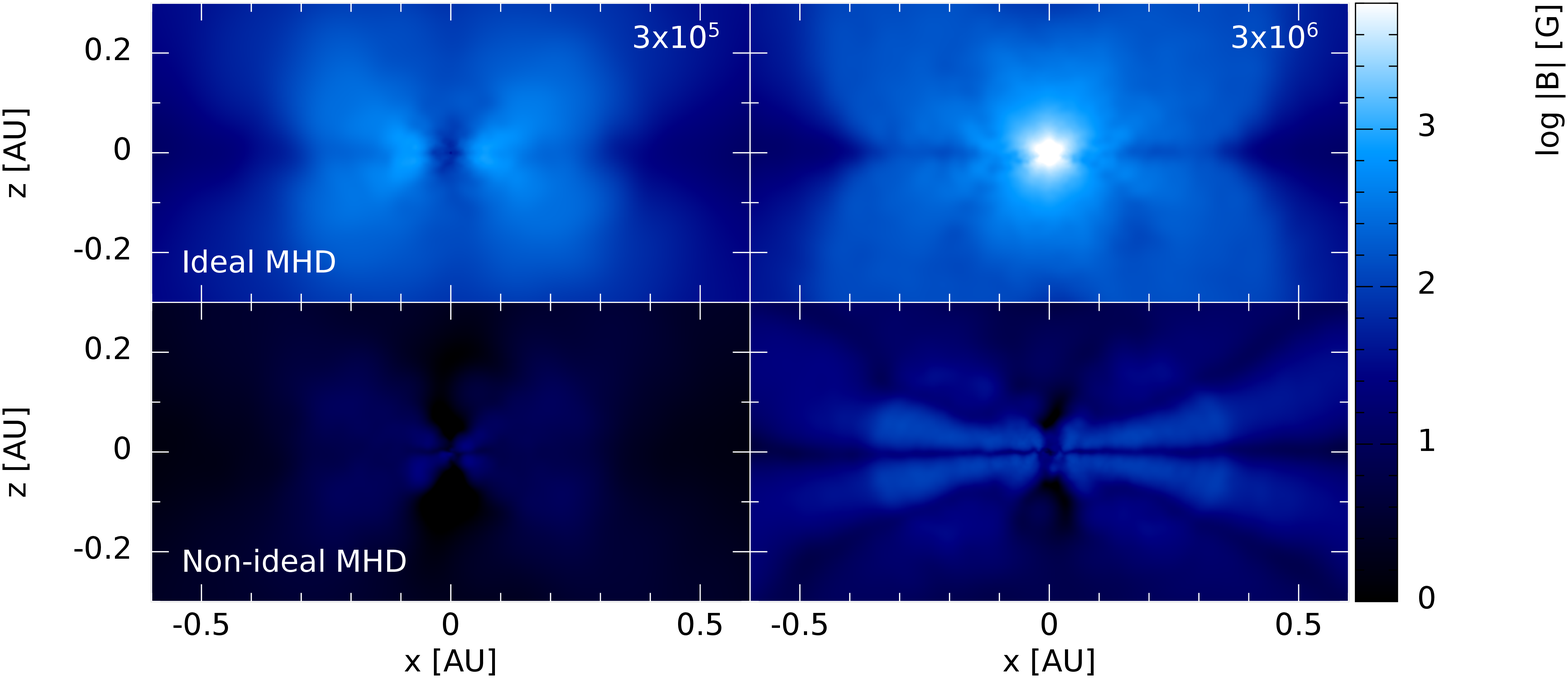}
\caption{Slices of magnetic field strength through the stellar core perpendicular to the rotation axis at \dtsc{0.5} for two different resolutions.  The structure of the ideal MHD model (top) is strongly dependent on the resolution, with different profiles of the central magnetic field strengths.  The non-ideal MHD model is also dependent on the resolution, with the magnetic field strength approximately 10 times weaker in the lower resolution model; both non-ideal MHD models, however, have qualitatively similar profiles, with the strongest magnetic field located in the disc around the stellar core which has a weaker field strength.}
\label{fig:B:res}
\end{figure*}

Artificial resistivity is second-order dependent on the resolution, thus increasing the resolution will decrease the importance of the artificial terms.  To illustrate the effect of numerical resolution, the first and third columns in Fig.~\ref{fig:slice:B:fhc} compare the magnetic field strength in a horizontal slice at two different resolutions during the first core phase.  The rows correspond to increasing maximum densities (top to bottom).

During the first core phase of the non-ideal MHD models, the maximum magnetic field strength is located in a torus near the edge of the first core.  Although physical resistivity is greater than artificial resistivity, the magnetic torus shows a dependence on both resolution and the choice of artificial resistivity algorithm.  Our previous studies \citep{WurPriBat2016,WurBatPri2018} used the algorithm developed by \citet{TriPri2013} (hereafter TP13); this method applies strong resistivity at strong magnetic field gradients.  This study, and \citet{WurPriBat2017,WurBatPri2018ion}, uses the artificial resistivity algorithm first presented by \citet{Phantom2017} (hereafter P18), which applies resistivity at strong velocity gradients.  A comparison of these two methods was presented in \citet{WurBatPriTri2017}, which showed the TP13 resistivity was generally more resistive than the P18 resistivity. Using TP13 (central column of Fig.~\ref{fig:slice:B:fhc}), the magnetic field is diffused out of the torus and into the centre of the first core, similar to the effect of decreasing the resolution with the P18 resistivity.  The magnetic torus remains strong when using P18, with the maximum magnetic field strength remaining in the torus rather than equilibrating between the centre and the surface of the first core.

Fig.~\ref{fig:BVrho:res} shows the evolution of the magnetic field with density after the formation of the first core for resolutions of $3\times10^5$ and $3\times10^6$ particles in the initial sphere for both our ideal and non-ideal MHD models.  During the final part of the first core phase and into the second collapse phase (\rhoxge{-9}), the low resolution model yields lower magnetic field strengths when using ideal MHD (top row).  By \dtsc{0.5}, the central magnetic field strengths differ by a factor of \sm5000, and the maximum magnetic field strengths differ by \sm100.  \citet{BatTriPri2014} showed in their Appendix A that the evolution of the magnetic field in ideal MHD is approximately converged between 3 and 10 million particles in the initial sphere.

As expected, the evolution of the magnetic field in the non-ideal MHD model (bottom row of Fig.~\ref{fig:BVrho:res}) is less dependent on the resolution.  In both models, the maximum magnetic field resides in the torus during the second collapse phase, which is necessarily better defined at higher resolution.  The higher resolution simulation results in a better defined magnetic wall where the magnetic field piles up, hence the jump of the maximum magnetic field strength (bottom right panel).  At lower resolutions, the artificial resistivity diffuses the magnetic wall, preventing the strong pile-up observed at higher resolutions.  This diffusion also results in the torus having a magnetic field strength that is only slightly larger than the central value.

At both resolutions, both the central and maximum magnetic field strengths at \dtsc{0.5} agree within a factor of \sm2.  Although there is some fluctuation in the magnetic field strengths, both models have central magnetic fields that are well below that required for fossil fields, indicating that our conclusions are independent of the specific artificial resistivity algorithm.

Fig.~\ref{fig:B:res} shows the magnetic field strength in a vertical slice through the stellar core at \dtsc{0.5} at two different resolutions (left and right; both use the P18 resistivity).  The differences between the two ideal MHD simulations is stark, with a strong magnetic field strength in the centre of the core for the fiducial resolution (right), whereas the magnetic field strength has been diffused out of the core in the low resolution model (left).  In both non-ideal models (bottom row), the magnetic field has been diffused out of the core and resides in the disc; this disc is smaller in the low resolution model, and its maximum strength is \sm10 times lower.  Both models show similar central magnetic field strengths (see Fig.~\ref{fig:BVrho:res}).

Computational resources currently prevent us from performing a non-ideal MHD simulation at $3\times10^7$ particles, thus we cannot conclusively show convergence.  However, the similarities between these two resolutions suggest that the weak central magnetic field in the stellar core is real.

%--------------------------------------------------------------------------------
\section{Summary and conclusion}
\label{sec:conc}
In this study, we modelled the collapse of a molecular cloud core through the first and stellar core phases in a strongly magnetized medium.  In the ideal MHD model, the maximum magnetic field strength was coincident with the maximum density, and grew to strengths a few orders of magnitude higher than observed for young stars; thus ideal MHD is a poor approximation when modelling star formation.  

In the non-ideal MHD model with the canonical cosmic ray ionization rate of \zetaeq{-17}, the maximum and central magnetic field strengths are no longer coincident during the first hydrostatic core phase, with the maximum magnetic field strength lying in a `magnetic wall' 1-3~au from the centre of the core.  Shortly after the formation of the stellar core the maximum magnetic field strength in the magnetic wall is $B_\text{max}\sim900$~G, while the central magnetic field strength reaches only $B_\text{cen}\sim$240~G.  Neither increasing the cosmic ray ionization rate by a factor of 10, nor switching to a more resistive artificial resistivity algorithm caused the central magnetic field strength to increase. 

Therefore, when a self-consistent treatment of non-ideal MHD is included in star formation simulations, the magnetic fields implanted in the stellar cores are lower than the kG magnetic fields that are required to provide the observed field strengths of young stars.  Since our results are sensitive to resolution, we cannot make a definitive conclusion, however, our results suggest that magnetic fields of low-mass stars cannot be fossil in origin, but must be generated through a subsequent dynamo process.

%--------------------------------------------------------------------------------
\section*{Acknowledgements}

We would like to thank the referee for useful and insightful comments that improved the quality of this manuscript.
JW and MRB acknowledge support from the European Research Council under the European Community's Seventh Framework Programme (FP7/2007- 2013 grant agreement no. 339248).  DJP received funding via Australian Research Council grants FT130100034, DP130102078 and DP180104235.  The calculations for this paper were performed on the DiRAC Complexity machine, jointly funded by STFC and the Large Facilities Capital Fund of BIS  (STFC grants ST/K000373/1, ST/K0003259/1 and ST/M006948/1), and the University of Exeter Supercomputer, a DiRAC Facility jointly funded by STFC, the Large Facilities Capital Fund of BIS, and the University of Exeter.
%\texttt{These simulations were run on both Zen and Complexity}.
The research data supporting this publication and \citet{WurBatPri2018ri} are openly available from the University of Exeter's institutional repository at: https://doi.org/10.24378/exe.607.
Several figures were made using \textsc{splash} \citep{Price2007}.  

\bibliography{fossil.bib}
%--------------------------------------------------------------------------------

\label{lastpage}
\end{document}